\documentclass[12pt,a4paper]{article}
\usepackage{graphicx}
\usepackage{soul}
\usepackage{epstopdf, epsfig,lmodern,pdftexcmds}
\usepackage{amsmath,tikz,commath,color} 
\usepackage{fullpage}

\usepackage{natbib}
\usepackage[title]{appendix}

\DeclareMathAlphabet{\mathsfbi}{OT1}{\sfdefault}{bx}{sl}
\DeclareMathVersion{sfletters}
\SetSymbolFont{letters}{sfletters}{OML}{ntxsfmi}{b}{it}

\newcommand{\bs}{\boldsymbol}
\newcommand{\mbi}{\mathsfbi}

\title{Viscous propulsion in active transversely-isotropic media}

\author{G. Cupples,$\!^1$ R. J. Dyson$^1$ and D. J. Smith$^{1,2,3}$\thanks{Email: D.J.Smith.2@bham.ac.uk. $^1$School of Mathematics and $^2$Institute for Metabolism and Systems Research, University of Birmingham, B15 2TT, U.K. $^3$Centre for Human Reproductive Science, Birmingham Women's NHS Foundation Trust, Birmingham, B15 2TG, U.K.}}

\begin{document}

\maketitle

\begin{abstract}
Taylor's swimming sheet is a classical model of microscale propulsion and pumping. Many biological fluids and substances are fibrous, having a preferred direction in their microstructure; for example cervical mucus is formed of polymer molecules which create an oriented fibrous network. Moreover, suspensions of elongated motile cells produce a form of active oriented matter. To understand how these effects modify viscous propulsion, we extend Taylor's classical model of small-amplitude zero-Reynolds-number propulsion of a `swimming sheet' via the transversely-isotropic fluid model of Ericksen, which is linear in strain rate and possesses a distinguished direction. The energetic costs of swimming are significantly altered by all rheological parameters and the initial fibre angle. Propulsion in a passive transversely-isotropic fluid produces an enhanced mean rate of working, independent of the initial fibre orientation, with an approximately linear dependence of energetic cost on the extensional and shear enhancements to the viscosity caused by fibres. In this regime the mean swimming velocity is unchanged from the Newtonian case. The effect of the constant term in Ericksen's model for the stress, which can be identified as a fibre tension or alternatively a stresslet characterising an active fluid, is also considered. This stress introduces an angular dependence and dramatically changes the streamlines and flow field; fibres aligned with the swimming direction increase the energetic demands of the sheet. The constant fibre stress may result in a reversal of the mean swimming velocity and a negative mean rate of working if sufficiently large relative to the other rheological parameters.
\end{abstract}

\section{Introduction}
Large organisms propel themselves through a fluid by utilising the inertia of the surrounding fluid. For very small organisms and cells swimming at low Reynolds numbers, inertial propulsion is not possible \citep{fauci2006biofluidmechanics, lauga2009hydrodynamics}; time-reversible kinematics result in no net displacement for the small body. G.I.\ Taylor's `swimming sheet' is one of the classical models of zero-Reynolds-number swimming; time-reversal symmetry is broken by the wave direction.

Many of the biological fluids in which these cells and organisms swim are non-Newtonian, hence modelling swimming in such fluids is of interest. The present study is motivated by the fibrous nature of many biological media, for example the cervical mucus encountered by the spermatozoa of many internally-fertilising species and active suspensions of elongated cells. Throughout the menstrual cycle, the rheology of cervical mucus changes due to hormonally-induced variations in hydration and associated changes in the glycofilament mucin structure. During ovulation these fibres form a parallel network (figure~\ref{fig:CervicalAlign}), and sperm migration occurs through this glycofilament structure \citep{chretien1982sperm, ceric2005ultrastructure}. It is therefore of great interest to determine how Stokesian swimming is modified by the presence of an aligned fibrous network.

\begin{figure}
\centering
\caption{Parallel filament mesh in cervical mucus during the time of ovulation. Bar = $10~\mu$m. Republished with permission of Oxford University Press, from `Ultrastructure of the Human Periovulatory Cervical Mucus', F. Ceric et al, 54 (5), 2005; permission conveyed through Copyright Clearance Center, Inc.}
\label{fig:CervicalAlign}
\end{figure}

\begin{figure}
\centering
\includegraphics[trim=1cm 1cm 1cm 0, clip, scale=0.8]{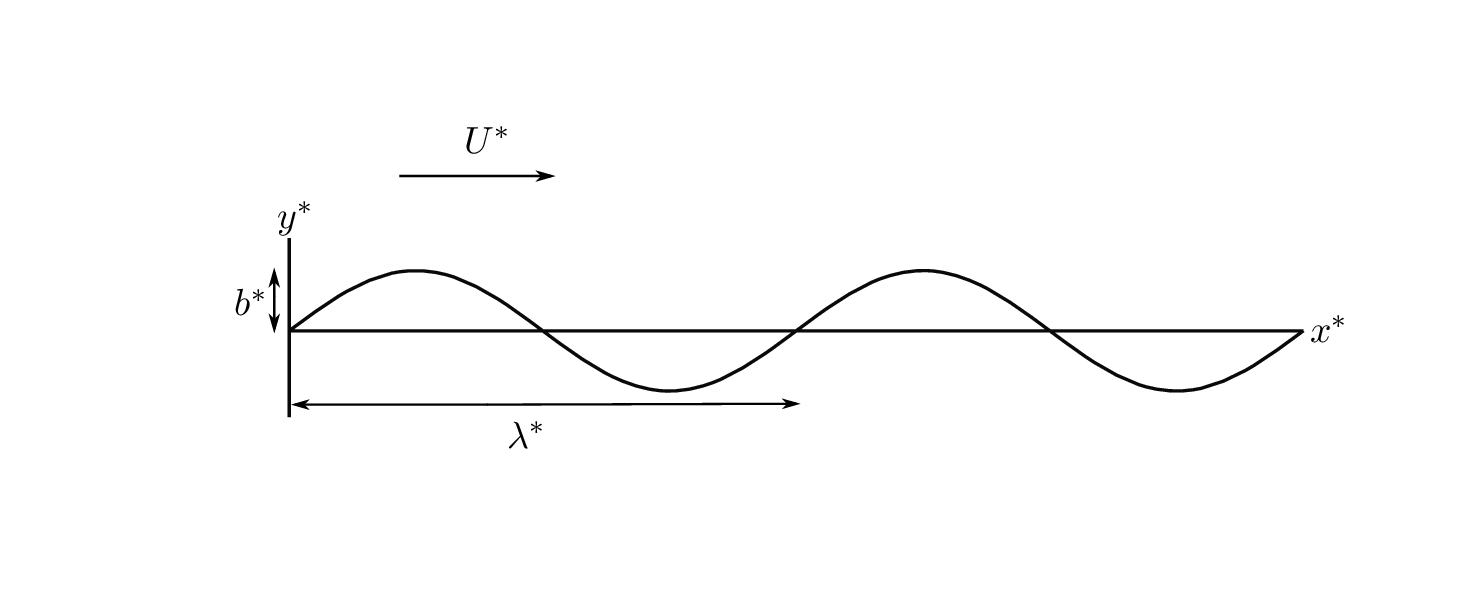}
\caption{A schematic of Taylor's swimming sheet in a Newtonian fluid. Working in a frame of reference in which the sheet is stationary, $b^*$ is the amplitude, $\lambda^*=2\pi/k^*$ is the wavelength and $k^*$
is the wavenumber. The flow at infinity in the $x^*$-direction is $U^*$. A travelling wave traverses the sheet with speed $c^*=\omega^*/k^*$ where $\omega^*$ is the angular velocity.}
\label{fig:taylorSchematic}
\end{figure}

\citeauthor{taylor1951swimming}'s pioneering study of Stokesian swimming consists of an infinite sheet undergoing waves of lateral displacement (figure~\ref{fig:taylorSchematic}). This model was formulated as the far-field Stokes flow produced by a swimming motion given by a small amplitude sinusoidal wave, and the associated mean rate of working was calculated as a measure of the energetic cost of swimming. Subsequent studies included a 3D model of a waving cylindrical tail \citep{taylor1952}, investigations by other authors into larger amplitude motion \citep{drummond1966propulsion} and more recently the unsteady Stokes flow problem \citep{pak2010transient}.

Generalising Taylor's model to non-Newtonian fluids has been an area of significant interest, for detailed review see \citet{lauga2009hydrodynamics}. \citet{chaudhury1979swimming} initially extended the model to incorporate viscoelastic fluids; it was found that the properties of the fluid leads to an increased steady swimming velocity for lower Reynolds numbers. This problem was reconsidered more recently by \citet{lauga2007propulsion}, who deduced that the mean swimming velocity in a nonlinear viscoelastic fluid is reduced relative to that in a Newtonian fluid, in certain cases the swimming direction is reversed \citep{fu2007theory}, also see \citep{fu2009swimming, teran2010viscoelastic}. \citet{velez2013waving} found propulsion in shear-thinning fluids to be more efficient than in Newtonian or shear-thickening fluids. \citet{riley2014} modelled active propulsion with fluid-structure interaction, and in a subsequent study deduced that for multiple travelling waves, the mean swimming velocity of the sheet is enhanced \citep{riley2015small}. Further to this, \citet{krieger2014locomotion, krieger2015minimal, krieger2015microscale} considered how liquid crystals affect the swimming of micro-organisms. Steady state and start-up models for hexatic liquid crystals were considered along with a nematic steady state model.

A transversely-isotropic fluid exhibits a (perhaps spatially and temporally varying) preferred direction, and has been used to model fibre-reinforced fluids.
Previous applications include the mechanical behaviour of collagen gel, the growth of plant root cell walls, suspensions of biomolecules and a multiphase model of extracellular matrix \citep{green2008extensional, dyson2010fibre, holloway2015couette, dyson2015investigation}. These models comprise a modified constitutive equation describing a viscous fluid with suspended aligned fibres and an expression for the evolution of fibre orientation. A transversely-isotropic fluid also provides a model of `active' suspensions of elongated swimmers \citep{holloway2016compare}.

In this study we consider swimming in transversely-isotropic fluids via the constitutive law of \citet{ericksen1960ti}. In Section \ref{sec:governingeqns} we introduce the governing equations associated with transversely-isotropic fluids. The problem is solved in Section \ref{sec:solutionmethod}, solving for the leading order velocity field, mean swimming velocity at next order and mean rate of working. The dependence of these quantities on the rheological parameters is explored in Section \ref{sec:Results}, and discussed in Section~\ref{sec:conc}.

\section{Governing equations} \label{sec:governingeqns}
The incompressibility and generalised  Navier-Stokes equations are
\begin{eqnarray}
\nabla^* \cdot \mathbf{u}^*&=&0,\label{incomp} \\
\rho^*\left(\frac{\partial \mathbf{u}^*}{\partial t^*}+(\mathbf{u}^*\cdot\nabla^*)\mathbf{u}^*\right)&=&\nabla^* \cdot \boldsymbol{\sigma}^*\label{nse},
\end{eqnarray}
where $\boldsymbol{u}^*=(u^*,v^*)$ is velocity, $\rho^*$ is density and $\boldsymbol{\sigma}^*$ is the stress tensor. We work in a 2D Cartesian coordinate system $(x^*,y^*)$; asterisk notation represents dimensional variables. A constitutive law is also required for $\boldsymbol{\sigma}^*$, which we prescribe in the next section.

\subsection{Transversely-isotropic stress tensor} \label{TIstress}
\citeauthor{ericksen1960ti}'s (1960) model consists of a stress tensor which is linear in strain rate, and depends on a unit vector \(\boldsymbol{a}\) describing the fibre orientation; this model takes the simplest form that satisfies the required invariances,
\begin{equation} \label{stress}
\sigma_{ij}^*=-p^*\delta_{ij}+2\mu^* e_{ij}^*+\mu_1^*a_ia_j+\mu_2^*a_ia_ja_ka_le_{kl}^*+2\mu_3^*(a_la_ie_{lj}^*+a_ma_je_{im}^*).
\end{equation}
We define $p^*$ as the pressure, $\delta_{ij}$ as the Kronecker delta function and $e_{ij}^*=\frac{1}{2}\left(\frac{\partial u_i^*}{\partial x_j^*}+\frac{\partial u_j^*}{\partial x_i^*}\right)$ as the rate-of-strain tensor \citep{ericksen1960ti}.

By considering simple flows with a uniform director field, we may interpret the rheological parameters as follows: by setting $\mu_1^*=\mu_2^*=\mu_3^*=0$, the stress tensor for an incompressible Newtonian fluid remains, with `matrix viscosity' $\mu^*$ \citep{holloway2015couette}. The term with $\mu_1^*$ has no dependence on velocity, suggesting that $\mu_1^*$ relates to a tension in the fibre direction \citep{dyson2010fibre}. This term can also be related to the stresslet-type active behaviour of fibres in a perfectly aligned active fluid \citep{holloway2016compare}. We will therefore refer to this quantity as the \emph{active parameter}. This term can be taken as a simple model for suspensions of self-propelling microscopic bodies such as bacteria or active gels of molecular motor proteins. The viscosity associated with extensional flow parallel to the fibre direction is $\mu_{||}^*=\mu^*+(\mu_2^*+4\mu_3^*)/2$, the viscosity associated with the flow orthogonal to the fibre direction is $\mu_{\perp}^*=\mu^*$ and the viscosity of shear flow in the fibre direction is $\mu_{s}^*=\mu^*+\mu_3^*$ \citep{dyson2010fibre}.
Since $\mu_2^*$ only has an impact on extensional viscosity parallel to the fibre direction, $\mu_\parallel^*$, it is termed the \emph{anisotropic extensional viscosity}. The parameter $\mu_3^*$ distinguishes $\mu_{\perp}^*$ from $\mu_s^*$ and so is labelled the \emph{anisotropic shear viscosity}; this parameter represents the difference between shear viscosities parallel and perpendicular to the fibre direction \citep{green2008extensional, dyson2010fibre, holloway2015couette}.

\begin{figure}
\centering
\includegraphics[trim=1cm 1cm 0cm 0.7cm,clip]{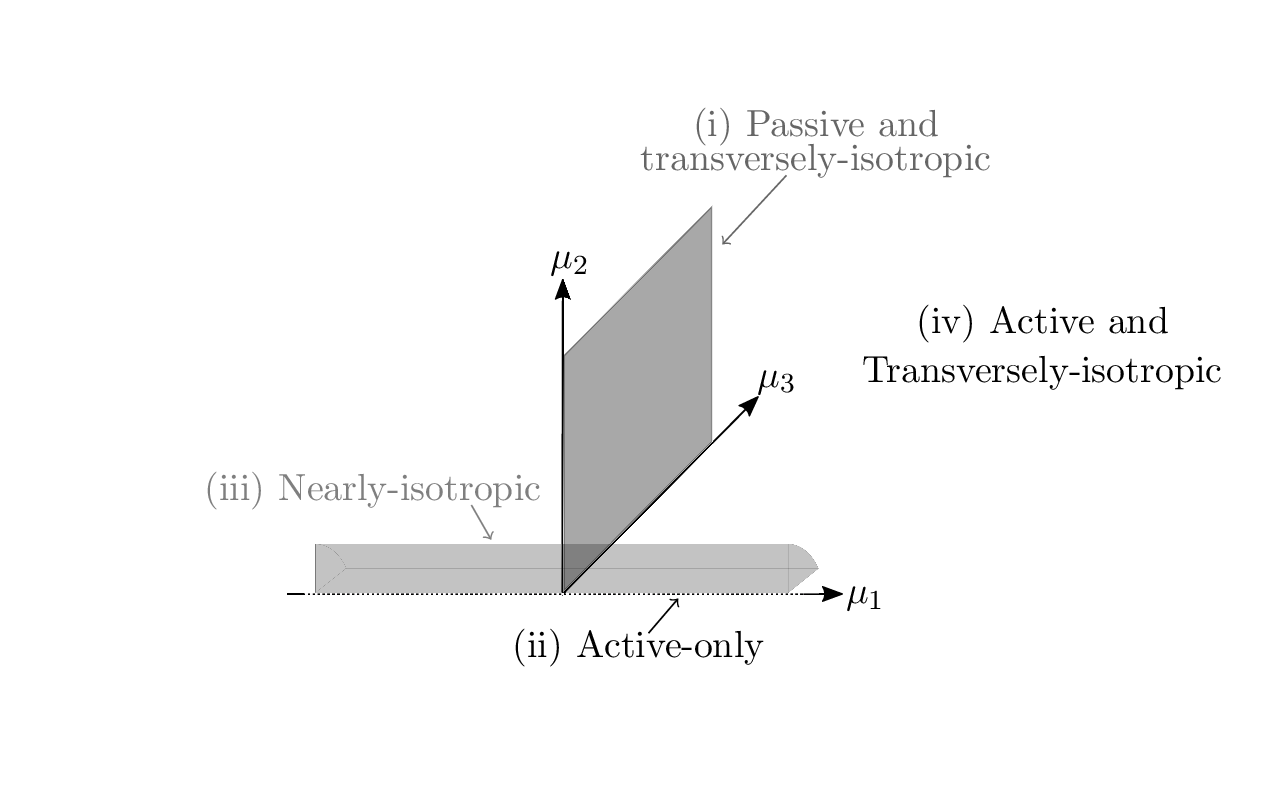}
\caption{Regimes of interest in parameter space. (i) The plane $\mu_1=0$ is the passive transversely-isotropic regime, (ii) the dashed line represents the active-only regime where $\mu_2=\mu_3=0$, (iii) the quarter cylinder is the nearly-isotropic regime where neither $\mu_2$ nor $\mu_3$ are large and (iv) the remaining region is the fully active and transversely-isotropic regime.}
\label{fig:Regions}
\end{figure}

\subsection{Fibre evolution equation} \label{kincond}

A fibre evolution equation describes the evolution of fibre orientation with time. We use the form given by \citet{green2008extensional},
\begin{equation} \label{kc}
\frac{\partial \boldsymbol{a}}{\partial t^*} + \boldsymbol{u}^* \cdot \nabla^*\boldsymbol{a} + [\boldsymbol{a} \cdot (\boldsymbol{a} \cdot \nabla^* \boldsymbol{u}^*)]\boldsymbol{a}=\boldsymbol{a}\cdot\nabla^*\boldsymbol{u}^*,
\end{equation}
which corresponds to a specific case of \citeauthor{ericksen1960ti}'s (\citeyear{ericksen1960ti}) equation in the long-fibre limit. Note that $|\bs{a}|=1$ and thus the model only considers local alignment of fibres and not their length. This gives a generalised form of Jeffery's treatment for long ellipsoidal particles aligning with flow \citep{jeffery1922motion, dyson2010fibre}.

Since $\bs{a}\cdot\bs{a}=1$, the component of equation \eqref{kc} in the $\bs{a}$-direction is automatically satisfied. The orthogonal component of equation \eqref{kc} is
\begin{equation} \label{kc1}
\bs{a}^\perp\cdot\left[\frac{\partial \boldsymbol{a}}{\partial t^*} + \boldsymbol{u}^* \cdot \nabla^*\boldsymbol{a} + [\boldsymbol{a} \cdot (\boldsymbol{a} \cdot \nabla^* \boldsymbol{u}^*)]\boldsymbol{a}-\boldsymbol{a}\cdot\nabla^*\boldsymbol{u}^*\right]=0,
\end{equation}
where $\bs{a}^\perp$ is a unit vector perpendicular to $\bs{a}$.

\subsection{Boundary conditions} \label{bc's}
We work in a frame of reference moving with the swimmer in the \(x^*\)-direction; the horizontal flow as \(y^*\rightarrow\infty\) therefore gives the mean swimming velocity. No-slip conditions on the sheet, representing a travelling wave with speed $c^*=\omega^*/k^*$, are thus
\begin{equation}  \label{bcdim}
u^* = 0, \quad v^* = -\omega^* b^* \cos(k^*x^*-\omega^* t^*), \quad \mbox{on } y^* = y_s^* = b^* \sin(k^*x^*-\omega^* t^*).
\end{equation}
The parameter $b^*$ is amplitude, $k^*$ is wave number, $y_s^*$ is the equation of the sheet surface and $\lambda^* = 2 \pi/k^*$ is wavelength. The velocity must remain bounded as $y^* \to \infty$.

\subsection{Non-dimensionalisation} \label{nondim}
The model is non-dimensionalised as follows:
\begin{equation}
\boldsymbol{u^*}=\frac{\omega^*}{k^*}\boldsymbol{u}, \quad {\bs x}^*=\frac{{\bs x}}{k^*}, \quad t^*=\frac{t}{\omega^*}, \quad p^*=\omega^*\mu^* p, \quad \bs{\sigma}^*=\omega^*\mu^*\bs{\sigma}. \label{scalings}
\end{equation}
The continuity and fibre evolution equations are unchanged. For microscopic swimmers, the Reynolds number, $Re=\rho^*\omega^*/{k^*}^2\mu^*$, is much less than one so we neglect inertial terms. The resulting system of partial differential equations is therefore
\begin{eqnarray}
\nabla\cdot\bs{u}&=&0, \label{momnondim} \\
\nabla\cdot\boldsymbol{\sigma}&=&\bs{0}, \label{cauchy} \\
\bs{a}^\perp\cdot\left[\frac{\partial \boldsymbol{a}}{\partial t} + \boldsymbol{u} \cdot \nabla\boldsymbol{a} + [\boldsymbol{a} \cdot (\boldsymbol{a} \cdot \nabla \boldsymbol{u})]\boldsymbol{a}-\boldsymbol{a}\cdot\nabla\boldsymbol{u}\right]&=&0, \label{kcnondim}
\end{eqnarray}
where
\begin{equation} \label{stressDL}
\sigma_{ij}=-p\delta_{ij}+2e_{ij}+\mu_1a_ia_j+\mu_2a_ia_ja_ka_le_{kl}+2\mu_3(a_la_ie_{lj}+a_ma_je_{im}),
\end{equation}
with dimensionless groups,
\begin{equation}
\mu_1=\frac{\mu_1^*}{\mu^*\omega^*}, \qquad \mu_2=\frac{\mu_2^*}{\mu^*}, \qquad  \mu_3=\frac{\mu_3^*}{\mu^*}.
\end{equation}
The boundary conditions \eqref{bcdim} become
\begin{equation} \label{dimlessbcs}
u = 0, \quad v = -\varepsilon\cos(x-t), \quad \mbox{on} \quad y=y_s=\varepsilon \sin(x-t),
\end{equation}
where $\varepsilon = k^* b^* \ll 1$. Also, $u$ and $v$ must remain bounded as $y\rightarrow\infty$.

Four regimes in parameter space, depicted in figure \ref{fig:Regions}, will be considered in our results: (i) a passive transversely-isotropic fluid, occurring when $\mu_1=0$, (ii) an active fluid ($\mu_1$ non-zero) with $\mu_2=\mu_3=0$, (iii) a nearly-isotropic regime, where all parameters take values up to $5$, and (iv) the regime where at least one of $\mu_1$, $\mu_2$ and $\mu_3$ are  much larger than one. Note that $\mu_1$ may be positive or negative, representing active `puller' or `pusher' behaviour respectively \citep{saintillan2010dilute}.

\section{Asymptotic solution} \label{sec:solutionmethod}

\begin{figure}
\centering
\includegraphics[trim=0.6cm 0cm 0cm 0.3cm,clip]{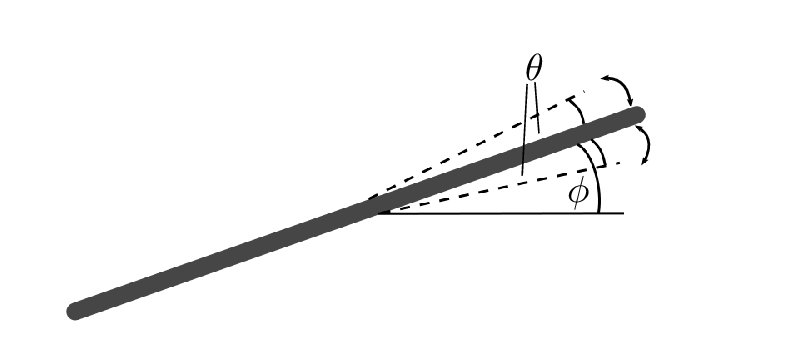}
\caption{A schematic showing the initial uniform orientation angle $\phi$ and the small perturbation away from this angle, $\theta$.}
\label{fig:Fibre}
\end{figure}

\subsection{Stream function formulation}

To determine the effect of fibres on the mean swimming velocity we consider an initially spatially-uniform fibre angle, $\phi$, aligned such that $\bs{a}(x,y,0)=(\cos\phi,\sin\phi)$. As the sheet swims this alignment will evolve, initially as a small perturbation $\theta(x,y,t)$ such that
\begin{eqnarray}\label{orientation vector}
\boldsymbol{a}&=&(\cos(\phi+\theta),\sin(\phi+\theta))\nonumber \\
			  &\approx &(\cos\phi-\theta\sin\phi, \sin\phi+\theta\cos\phi),
\end{eqnarray}
(see figure \ref{fig:Fibre}). The components of the stress tensor in terms of $\phi$ and $\theta$ are given in Appendix \ref{appA}. Taking the curl of equation \eqref{cauchy} eliminates pressure, reducing \eqref{cauchy} to a single equation. Since the flow is incompressible, we introduce a dimensionless stream function
\begin{equation}
u=\dpd{\psi}{y}, \quad v=-\dpd{\psi}{x},
\end{equation}
transforming equation \eqref{cauchy} to
\begin{eqnarray}
(1&+&\frac{\mu_2}{4}\sin^22\phi+\mu_3)\nabla^4\psi-\mu_1\left[2\sin2\phi\left(\theta\left(\dpd[2]{\theta}{y}-\dpd[2]{\theta}{x}\right)+\left(\dpd{\theta}{y}\right)^2-\left(\dpd{\theta}{x}\right)^2\right.\right. \nonumber \\
& & \quad \left.\left. + \dmd{\theta}{2}{x}{}{y}{}\right)+\cos2\phi\left(\dpd[2]{\theta}{x}-\dpd[2]{\theta}{y}+4\left(\dpd{\theta}{x}\dpd{\theta}{y}+\theta\dmd{\theta}{2}{x}{}{y}{}\right)\right)\right] \nonumber \\
&-&\mu_2\left[\sin4\phi\left(-\frac{\theta}{2}\left(\dpd[4]{\psi}{x}-3\dmd{\psi}{4}{x}{2}{y}{2}+\dpd[4]{\psi}{y}\right) +\frac{1}{2}\left(\dpd[2]{\theta}{x}-\dpd[2]{\theta}{y}\right)\left(\dpd[2]{\psi}{y}-\dpd[2]{\psi}{x}\right)\right.\right. \nonumber \\
& & \hspace{1.3cm} \left.\left.-\dpd{\theta}{y}\left(\dpd[3]{\psi}{y}-3\dmd{\psi}{3}{x}{2}{y}{}\right)+2\dmd{\theta}{2}{x}{}{y}{}\dmd{\psi}{2}{x}{}{y}{} +\dpd{\theta}{x}\left(3\dmd{\psi}{3}{x}{}{y}{2}-\dpd[3]{\psi}{x}\right)\right.\right. \nonumber \\
& & \hspace{1.1cm} \left.\left. +\theta\dmd{\psi}{4}{x}{2}{y}{2}+\frac{1}{2}\left(\dmd{\psi}{4}{x}{3}{y}{}-\dmd{\psi}{4}{x}{}{y}{3}\right)\right)+\cos4\phi\left(2\theta\left(\dmd{\psi}{4}{x}{3}{y}{}-\dmd{\psi}{4}{x}{}{y}{3}\right) \right.\right. \nonumber \\
& & \hspace{2cm}\left.\left. +\left(\dpd[2]{\theta}{x}-\dpd[2]{\theta}{y}\right)\dmd{\psi}{2}{x}{}{y}{}-\dpd{\theta}{x}\left(\dpd[3]{\psi}{y}-3\dmd{\psi}{3}{x}{2}{y}{}\right)\right.\right. \nonumber \\
& & \hspace{1cm} \left.\left. -\dpd{\theta}{y}\left(3\dmd{\psi}{3}{x}{}{y}{2}-\dpd[3]{\psi}{x}\right)-\dmd{\theta}{2}{x}{}{y}{}\left(\dpd[2]{\psi}{y}-\dpd[2]{\psi}{x}\right)-\dmd{\psi}{4}{x}{2}{y}{2}\right)\right]=0. \label{FullGov}
\end{eqnarray}

When $\mu_1=\mu_2=\mu_3=0$, equation \eqref{FullGov} reduces to the familiar biharmonic equation of Newtonian Stokes flow. The evolution equation \eqref{kcnondim} becomes
\begingroup
\addtolength{\jot}{0.7em}
\begin{eqnarray}
\dpd{\theta}{t}&+&\left[\dpd{\psi}{y}\dpd{\theta}{x}-\dpd{\psi}{x}\dpd{\theta}{y}\right] + (\sin^2\phi+\frac{\theta}{2}\sin2\phi)\dpd[2]{\psi}{y} \nonumber \\
& & \quad + (\sin2\phi+\theta\cos2\phi)\dmd{\psi}{2}{x}{}{y}{}+(\cos^2\phi-\frac{\theta}{2}\sin2\phi)\dpd[2]{\psi}{x} \nonumber \\
& & \quad + \theta\Bigg(\left[(\cos\phi-\theta\sin\phi)^2-(\sin\phi+\theta\cos\phi)^2\right]\dmd{\psi}{2}{x}{}{y}{}\nonumber \\
& & \hspace{1.7cm}+(\cos\phi-\theta\sin\phi)(\sin\phi+\theta\cos\phi)\left(\dpd[2]{\psi}{y}-\dpd[2]{\psi}{x}\right)\Bigg)=0. \label{FullKc}
\end{eqnarray}
\endgroup
The boundary conditions \eqref{dimlessbcs} become
\begin{equation}
\dpd{\psi}{y}=0, \quad \dpd{\psi}{x}=\varepsilon\cos(x-t) \quad \mbox{on}\, y=\varepsilon\sin(x-t),
\end{equation}
with $\psi$ having bounded first derivatives as $y \to \infty$.

\subsection{Perturbation expansion} \label{sec:PertExpand}

To apply the boundary conditions at $y=0$ rather than on the sheet, we make the small amplitude expansion
\begingroup
\addtolength{\jot}{0.7em}
\begin{eqnarray}
\left. \frac{\partial \psi}{\partial y} \right|_{y=0} + \left. \varepsilon \sin(x-t) \frac{\partial^2 \psi}{\partial y^2} \right|_{y=0} + ... &=&0, \label{Taylor expanded bc 1}\\
\left. \frac{\partial \psi}{\partial x} \right|_{y=0} + \left. \varepsilon \sin(x-t) \frac{\partial^2 \psi}{\partial y \partial x} \right|_{y=0} + ... &=&\varepsilon \cos(x-t).\label{Taylor expanded bc 2}
\end{eqnarray}
\endgroup

The velocity and fibre angle perturbations thus take the form
\begin{align}
\psi(x,y,t;\varepsilon) &= \varepsilon\psi_0(x,y,t) + \varepsilon^2 \psi_1(x,y,t)+ ...\label{PertExpansionPsi}\\
\theta(x,y,t;\varepsilon)&=\varepsilon\theta_0(x,y,t)+\varepsilon^2\theta_1(x,y,t)+...\label{PertExpansionTheta}
\end{align}
As in Taylor's analysis, the background flow ({\it i.e.} the sheet swimming velocity) will occur at order $\varepsilon^2$.

\subsection{Leading-order solution} \label{LOsol}
Substituting the expansions into the equations \eqref{FullGov} and \eqref{FullKc} and equating coefficients of powers of $\varepsilon$ yields the leading order partial differential equation.
At order $\varepsilon$, equation \eqref{FullGov} yields
\begin{eqnarray}
(1&+&\frac{\mu_2}{4}\sin^22\phi+\mu_3)\nabla^4\psi_0-\mu_1\left(2\sin2\phi\dmd{\theta_0}{2}{x}{}{y}{}+\cos2\phi\left(\dpd[2]{\theta_0}{x}-\dpd[2]{\theta_0}{y}\right)\right) \nonumber \\
&+&\mu_2\left(\cos4\phi\dmd{\psi_0}{4}{x}{2}{y}{2}+\frac{\sin4\phi}{2}\left(\dmd{\psi_0}{4}{x}{}{y}{3}-\dmd{\psi_0}{4}{x}{3}{y}{}\right)\right)=0,\label{AngleGov}
\end{eqnarray}
and equation \eqref{FullKc}
\begin{equation} \label{Anglekc}
\dpd{\theta_0}{t}+\sin2\phi\dmd{\psi_0}{2}{x}{}{y}{}+\cos^2\phi\dpd[2]{\psi_0}{x}+\sin^2\phi\dpd[2]{\psi_0}{y}=0.
\end{equation}
The boundary conditions \eqref{Taylor expanded bc 1} and \eqref{Taylor expanded bc 2} become
\begin{equation}\label{LObc}
\dpd{\psi_0}{y}=0, \quad \dpd{\psi_0}{x}=\cos(x-t), \quad \mbox{on } y = 0,
\end{equation}
combined with the requirement that the derivatives of \(\psi_0\) are bounded as \(y\rightarrow \infty\).

Equations \eqref{AngleGov} and \eqref{Anglekc} are solved with the ansatz,
\begin{eqnarray}
\psi_0&=&f_1(y)\cos(x-t)+f_2(y)\sin(x-t),\\
\theta_0&=&g_1(y)\cos(x-t)+g_2(y)\sin(x-t),
\end{eqnarray}
for some functions $f_1,\, f_2,\, g_1,\, g_2$. Comparing coefficients of sine and cosine leads to a system of four ordinary differential equations
\begingroup
\addtolength{\jot}{0.7em}
\begin{eqnarray}
(1+\frac{1}{4}\mu_2\sin^22\phi+&\mu_3)(f_1''''-2f_1''+f_1)+\mu_1(\cos2\phi\,(g_1+g_1'')-2\sin2\phi\, g_1')\nonumber \\
&+\mu_2(\frac{1}{2}\sin4\phi\,(f_2'''+f_2'')-\cos4\phi\, f_1'')=0,\label{govsin}\\
(1+\frac{1}{4}\mu_2\sin^22\phi+&\mu_3)(f_2''''-2f_2''+f_2)+\mu_1(\cos2\phi\,(g_2+g_2'')+2\sin2\phi\, g_1')\nonumber \\
&-\mu_2(\frac{1}{2}\sin4\phi\,(f_1'''+f_1'')+\cos4\phi\, f_2'')=0,\label{govcos}\\
& \hspace{0.3cm} g_1-\sin2\phi\, f_1'+\sin^2\!\phi\, f_2''-\cos^2\!\phi\, f_2=0,\label{kcsin}\\
& \hspace{0.3cm} g_2-\sin2\phi\, f_2'-\sin^2\!\phi\, f_1''+\cos^2\!\phi\, f_1=0,\label{kccos}
\end{eqnarray}
\endgroup
where prime denotes differentiation with respect to $y$.

Substituting equations \eqref{kcsin} and \eqref{kccos} into \eqref{govsin} and \eqref{govcos}, the system reduces to two ordinary differential equations for $f_1$ and $f_2$. Assuming a basis of solutions of the form
\begingroup
\addtolength{\jot}{0.7em}
\begin{equation}
\renewcommand{\arraystretch}{1.1}
\begin{pmatrix}
f_1\\
f_2
\end{pmatrix}=
\begin{pmatrix}
f_1^0\\
f_2^0
\end{pmatrix}e^{\lambda y},
\end{equation}
\endgroup
reduces the problem to the linear system,
\begingroup
\addtolength{\jot}{0.7em}
\begin{equation}
\renewcommand{\arraystretch}{1.1}
\mbi{L}
\begin{pmatrix}
f_1^0\\
f_2^0
\end{pmatrix}=
\begin{pmatrix}
0 \\
0
\end{pmatrix}
\quad \mbox{where} \quad
\mbi{L}=
\begin{pmatrix}
{L}_{11} & {L}_{12} \\
{L}_{21} & {L}_{22}
\end{pmatrix},
\end{equation}
\endgroup
where components of $\mbi{L}$ are given in Appendix \ref{appB}. Note that ${L}_{11}={L}_{22}$ and further that ${L}_{12}=-{L}_{21}$. For a non-trivial solution, the determinant of the matrix $\mbi{L}$ must be zero, yielding the equation
\begin{equation}
{L}_{11}^2+{L}_{12}^2=0\label{rootseq},
\end{equation}
hence
\begin{equation}
{L}_{11}=\pm i {L}_{12},
\end{equation}
and so
\begingroup
\addtolength{\jot}{0.7em}
\begin{eqnarray}
f_2^0 &=&- \frac{{L}_{11}}{{L}_{12}}f_1^0,\nonumber \\
&=&\mp i f_1^0.
\end{eqnarray}
\endgroup

Equation \eqref{rootseq} has eight complex roots, $\lambda_j$, four with positive real part and four with negative real part. Since the velocity must remain bounded as $y\rightarrow\infty$, we disregard the positive roots. The other four form two complex conjugate pairs,
\begin{eqnarray}
\lambda_1=\alpha_1+i\beta_1, &\qquad \lambda_3=\alpha_1-i\beta_1,\\
\lambda_2=\alpha_2+i\beta_2, &\qquad \lambda_4=\alpha_2-i\beta_2.
\end{eqnarray}
Note that $\lambda_j$ are known analytically, however they are not given here due to space constraints.

The solution form for $\psi_0$ is thus
\begin{equation}\label{streamfunctionLO}
\psi_0=\sum_{j=1}^4 \hat{A_j}(\cos(x-t)+\xi_j\sin(x-t))e^{\lambda_j y},
\end{equation}
where $\xi_j=-i$ for $j=1,2$ and $\xi_j=i$ for $j=3,4$. Assuming that the constants take the general form $\hat{A}_j=A_j+iB_j$ for $j=1,\,2,\,3,\,4$, boundary conditions \eqref{LObc} give
\begingroup
\addtolength{\jot}{0.7em}
\begin{eqnarray*}
A_1&=&\frac{\alpha_1\beta_2-\alpha_2\beta_1}{2((\alpha_1-\alpha_2)^2+(\beta_1-\beta_2)^2)}, \quad A_2=-A_1, \quad A_3=A_1, \quad A_4=-A_1, \\
B_1&=&\frac{\alpha_2^2-\alpha_1\alpha_2+\beta_2^2-\beta_1\beta_2}{2((\alpha_1-\alpha_2)^2+(\beta_1-\beta_2)^2)}, \quad B_3=-B_1, \\
B_2&=&\frac{\alpha_1^2-\alpha_1\alpha_2+\beta_1^2-\beta_1\beta_2}{2((\alpha_1-\alpha_2)^2+(\beta_1-\beta_2)^2)}, \quad B_4=-B_2.
\end{eqnarray*}
\endgroup
The fibre angle perturbation is then of the form
\begin{eqnarray}
\theta_0&=&\sum_{j=1}^4 \hat{A}_j\left[(\lambda_j\sin2\phi+\xi_j(-\lambda_j^2\sin^2\phi+\cos^2\phi))\cos(x-t)\right.\nonumber \\
& &\qquad \qquad \qquad \left. +(\xi_j\lambda_j\sin2\phi+\lambda_j^2\sin^2\phi-\cos^2\phi)\sin(x-t)\right]e^{\lambda_j y}. \label{theta0}
\end{eqnarray}
The change in the small perturbation to the orientation, $\theta_0$, is dependent on the initial orientation angle of the fibres along with their position.

\subsection{Order $\varepsilon^2$ solution and mean swimming velocity}
The mean swimming velocity is determined by the horizontal component of the flow as $y$ tends to infinity. The leading-order stream function, \eqref{streamfunctionLO}, tends to zero and hence the non-zero mean swimming velocity is determined at order $\varepsilon^2$
\begin{equation} \label{leadingswimvel}
U \sim \varepsilon^2 U_1=\lim_{y\rightarrow\infty}\varepsilon^2 \dpd{\psi_1}{y}.
\end{equation}
We neglect the oscillatory terms to determine the leading-order term in the expansion of mean swimming velocity, which we denote as $\overline{U}_1$. The bar notation represents an average over one time period.

At order $\varepsilon^2$, the boundary conditions \eqref{Taylor expanded bc 1} and \eqref{Taylor expanded bc 2} become
\begin{equation} \label{FObc}
\frac{\partial \psi_1}{\partial y}\bigg|_{y=0} +  \sin(x-t) \frac{\partial^2 \psi_0}{\partial y^2}\bigg|_{y=0} = 0, \quad \frac{\partial \psi_1}{\partial x}\bigg|_{y=0} + \sin(x-t) \frac{\partial^2 \psi_0}{\partial y \partial x}\bigg|_{y=0} = 0,
\end{equation}
and hence
\begingroup
\addtolength{\jot}{0.7em}
\begin{eqnarray}
\dpd{\psi_1}{y}\bigg|_{y=0}&=&\frac{1}{2}\left((\alpha_1\alpha_2-\beta_1\beta_2)(1-\cos 2(x-t))-(\alpha_1\beta_2-\alpha_2\beta_1)\sin 2(x-t)\right)\! ,\label{ep2bc1}\\
& &\hspace{1.8cm} \dpd{\psi_1}{x}\bigg|_{y=0}=0.\label{ep2bc2}
\end{eqnarray}
\endgroup

Because of the form of the boundary conditions, the ansatz is
\begin{equation}
\psi_1=\hat{f}_1(y)+\hat{f}_2(y)\cos 2(x-t)+\hat{f}_3(y)
\sin 2(x-t), \label{psi1}
\end{equation}
for some functions $\hat{f}_1,\, \hat{f}_2$ and $\hat{f}_3$. Substituting the solution form \eqref{psi1} into the order $\varepsilon^2$ expansion of \eqref{FullGov}, shown in full in Appendix \ref{appC}, and equating coefficients of non-oscillating terms, we have
\begin{equation}
\hat{f}_1''''=0,
\end{equation}
and hence
\begin{equation}
\hat{f}_1(y)=Ay^3+By^2+Cy+D.
\end{equation}
To ensure the velocity remains bounded as $y\rightarrow\infty$, we set $A=B=0$. From the boundary conditions \eqref{ep2bc1} and \eqref{ep2bc2} we find that $C=(\alpha_1\alpha_2-\beta_1\beta_2)/2$ and $D=0$, and hence
\begin{equation}
\hat{f}_1(y)=\frac{y}{2}(\alpha_1\alpha_2-\beta_1\beta_2).
\end{equation}
Differentiating $\hat{f}_1(y)$ with respect to $y$, the leading-order term in the expansion of mean swimming velocity is calculated as
\begin{equation}\label{backgroundflow}
\overline{U}_1=\frac{1}{2}(\alpha_1\alpha_2-\beta_1\beta_2).
\end{equation}

\subsection{Mean rate of working} \label{sec:energydiss}

To determine how Stokesian swimming is affected by transverse isotropy, the mean rate of working at order $\varepsilon$ is investigated, {\it i.e.}\ the rate of working per unit area of the sheet against viscous stress,  \(\varepsilon^2\overline{W}\) \citep{taylor1951swimming}. The mean value of this quantity is given by
\begin{equation}\label{workdissipationeqn}
\varepsilon^2\overline{W}=-\overline{\frac{\partial y_s}{\partial t}\sigma_{22}|_{y=0}},
\end{equation}
where $y_s$ is the equation of the sheet surface and $\sigma_{22}|_{y=0}$ is the normal stress evaluated on the sheet. The no-slip condition is $u=0$ on the sheet and hence $\partial u/\partial x=0$ and, via \eqref{momnondim}, $\partial{v}/\partial y=0$. In terms of the stream function, $\sigma_{22}$ is
\begin{equation}
\sigma_{22}=-p+\mu_1\sin^2\phi +(\mu_2\cos\phi\sin^3\phi+\mu_3\sin2\phi)\left(\dpd[2]{\psi}{y}-\dpd[2]{\psi}{x}\right).
\end{equation}

Solving equation \eqref{cauchy}, using the leading-order expression for the stream function \eqref{streamfunctionLO}, determines pressure. Noting that $\overline{\cos^2(x-t)}=1/2$, we obtain an expression for the leading order term in the expansion of mean rate of working,
\begin{eqnarray}
\overline{W}&=&-\frac{1}{16}\Big[\left(\alpha_1^2\alpha_2-\alpha_2\beta_1(\beta_1+2\beta_2)\right. \nonumber \\
& &\hspace{1.6cm} \left.+\alpha_1(\alpha_2^2-\beta_2(2\beta_1+\beta_2))\right)(8+\mu_2(1+\cos4\phi)+8\mu_3) \nonumber \\
& &\hspace{3cm}  + 4(\alpha_2\beta_1+\alpha_1\beta_2)\mu_2\sin4\phi\Big]. \label{WD}
\end{eqnarray}

\section{Results} \label{sec:Results}

The leading-order terms in the expansions of mean swimming velocity \eqref{backgroundflow}, mean rate of working \eqref{WD}, fibre perturbation \eqref{theta0}, velocities $u_0=\partial \psi_0/\partial y$, $v_0=-\partial\psi_0/\partial x$ and stream function \eqref{streamfunctionLO}, have been found analytically in terms of lengthy expressions for $\alpha_j,\, \beta_j$. The analytical results for the mean rate of working have been recreated numerically using finite differences and integration by the midpoint method. Each separate component has been verified along with the full solution. The solutions agreed to within a small degree of numerical error. For brevity we will refer to the time averages of the leading order terms in the expressions for swimming velocity and rate of working as the \emph{mean swimming velocity} and \emph{mean rate of working} respectively, and we will plot terms without the leading \(\varepsilon^2\) factors as defined by $\overline{U}_1,\, \overline{W}$ in equations~\eqref{leadingswimvel} and \eqref{workdissipationeqn}.

We now discuss the results in more detail. Four different flow regimes are considered (figure \ref{fig:Regions}): (i) a passive transversely isotropic fluid, occurring when $\mu_1=0$, (ii) an active fluid where $\mu_2=\mu_3=0$ and $\mu_1$ is non-zero, (iii) a nearly isotropic regime, where all parameters take values up to $5$ and (iv) the regime where at least one of $\mu_1,\, \mu_2$ and $\mu_3$ are much larger than one. A range of initial orientation angle $\phi$, between $0$ and $2\pi$ are considered for all regimes and the active parameter, $\mu_1$, is allowed to take both positive and negative values to account for `puller' and `pusher' active behaviour respectively. Note that since the fibres have no directionality, the regime $\phi=0$ to $\pi$ is identical to $\phi=\pi$ to $2\pi$.

\subsection{Regime (i): Effect of passive fibres on mean swimming velocity and rate of working}

\begin{figure}
\centering
\includegraphics[scale=0.63, trim=3.5cm 1.2cm 0.5cm 1cm,clip]{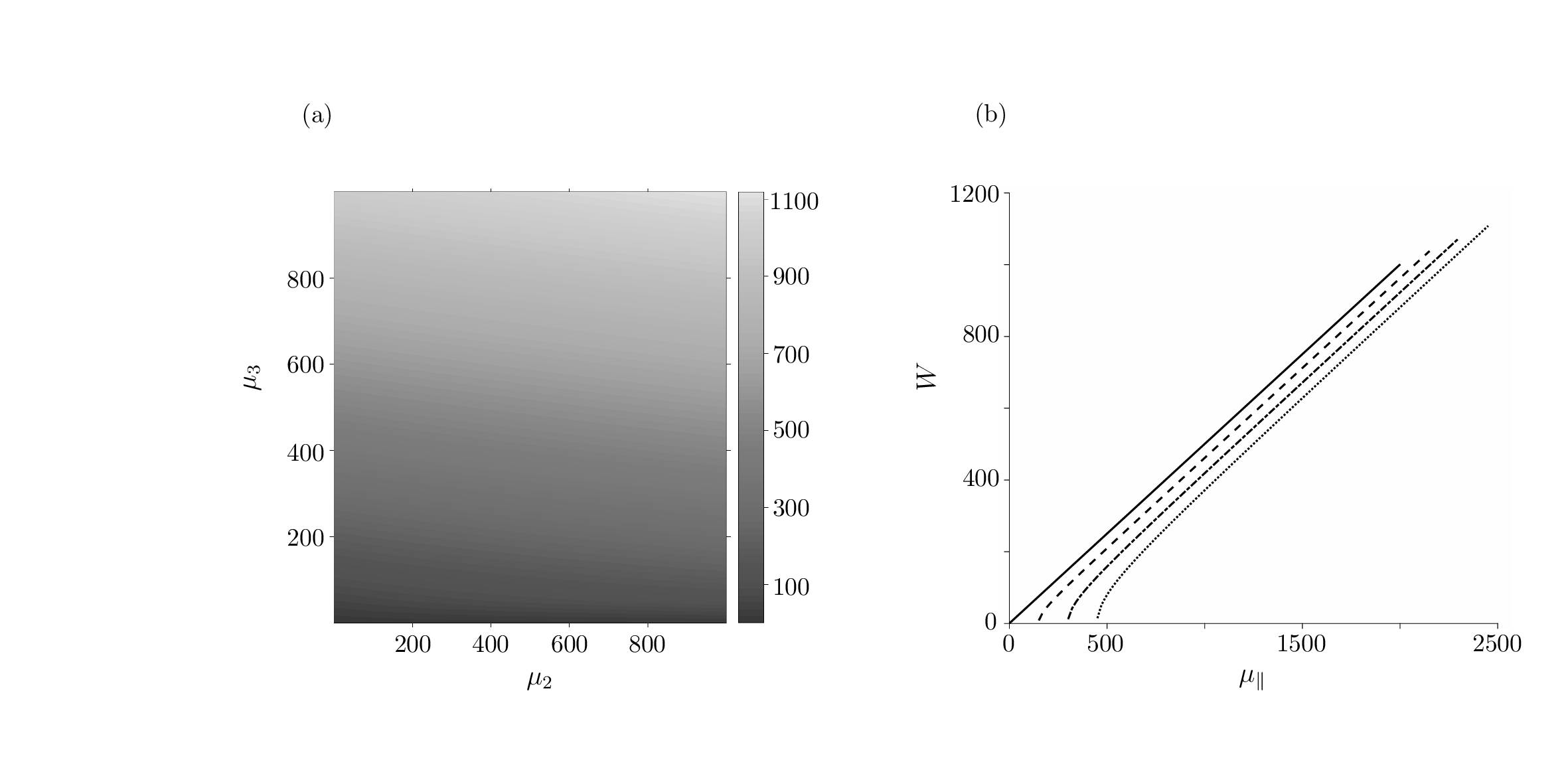}
\caption{Mean rate of working by the swimmer in a transversely-isotropic fluid where $\mu_1=0$. (a) Depicts mean rate of working for varying $\mu_2$ and $\mu_3$. This result is identical for all initial angles $\phi$. (b) Depicts how mean rate of working changes for increasing $\mu_{\parallel}$ for a range of $\mu_3$ and set $\mu_2$ values; $\mu_2=0$ (solid line), $\mu_2=300$ (dashed line), $\mu_2=600$ (dash-dotted line) and $\mu_2=900$ (dotted line), where the arrow denotes increasing $\mu_2$.}
\label{fig:NoMove}
\end{figure}

Passive fibres exert no shear-independent force and have no self-propulsive properties, hence the active parameter $\mu_1$ is set to zero. In this regime the mean swimming velocity takes on the Newtonian value, $\overline{U}_1=1/2$, and the mean rate of working is independent of the initial orientation angle $\phi$. In figure \ref{fig:NoMove} (a), the mean rate of working is always greater than or equal to the Newtonian case, $\overline{W}=1$. The increase in mean rate of working is linear throughout, with $\mu_3$ having a bigger impact than $\mu_2$. Figure \ref{fig:NoMove} (b) depicts the relationship between the mean rate of working and the parallel viscosity $\mu_{\parallel}=1+(\mu_2+4\mu_3)/2$, where each line represents a different $\mu_2$. The increase in mean rate of working with $\mu_\parallel$ is linear, apart from the case where $\mu_3$ is small and $\mu_2$ is large, with a large mean rate of working as $\mu_2$ increases.

\subsection{Regime (ii): Active-only effects on mean swimming velocity and rate of working}

\begin{figure}
\centering
\includegraphics[scale=0.7, trim=1.1cm 1.7cm 2.3cm 0.8cm,clip]{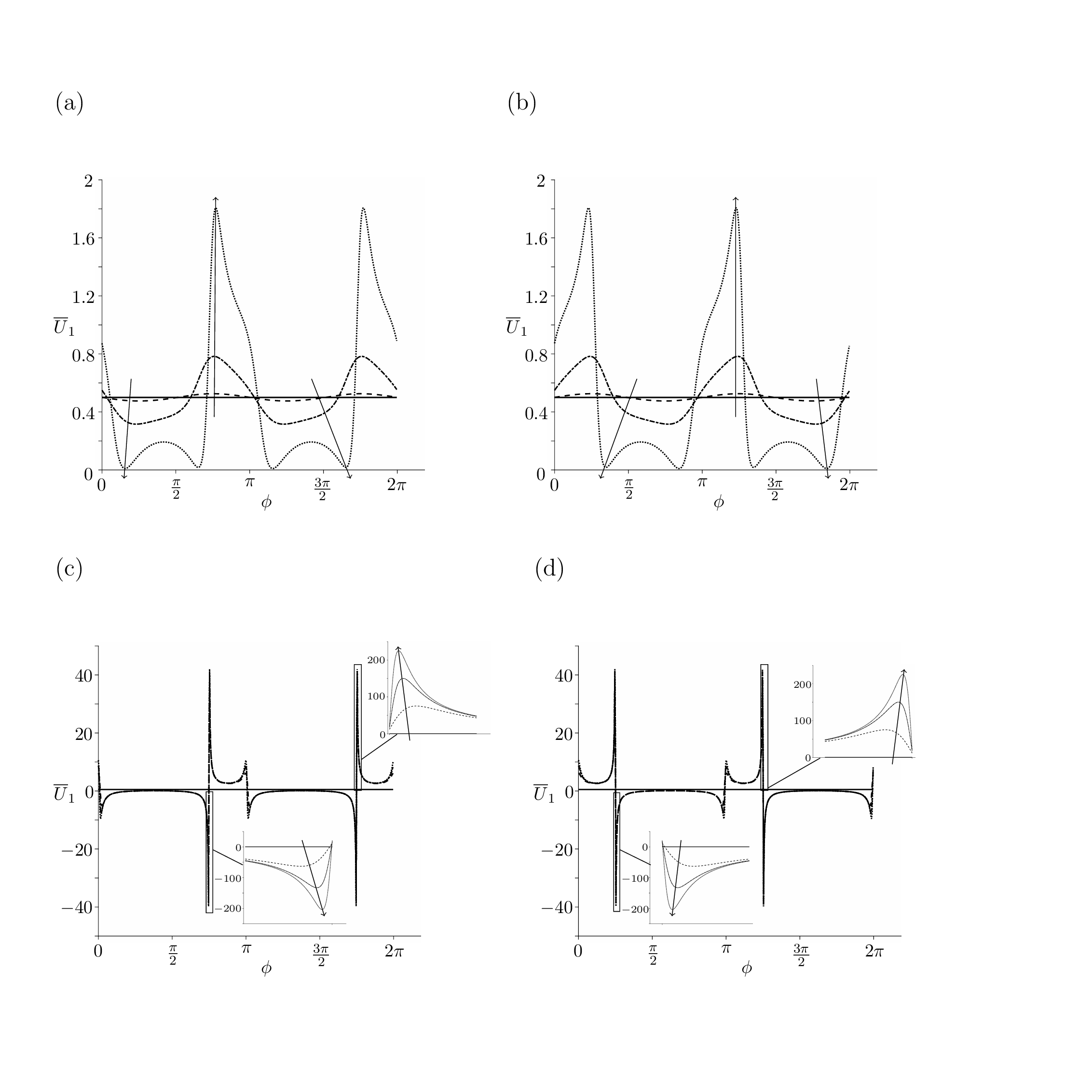}
\caption{Mean swimming velocity versus the initial orientation $\phi$, where $\mu_2=\mu_3=0$. (a) and (b) depict small positive and negative $\mu_1$ values: $0$ (solid line), $\pm 0.1$ (dashed line), $\pm 1$ (dash-dotted line) and $\pm 5$ (dotted line). (c) and (d) depict larger $\mu_1$ values; $0$ (solid line), $\pm 300$ (dashed line), $\pm 600$ (dash-dotted line), $\pm 900$ (dotted line), where the arrows denotes increasing $\mu_1$.}
\label{fig:ActiveOnlyU}
\end{figure}

\begin{figure}
\centering
\includegraphics[scale=0.7, trim=1cm 1.5cm 2cm 1cm,clip]{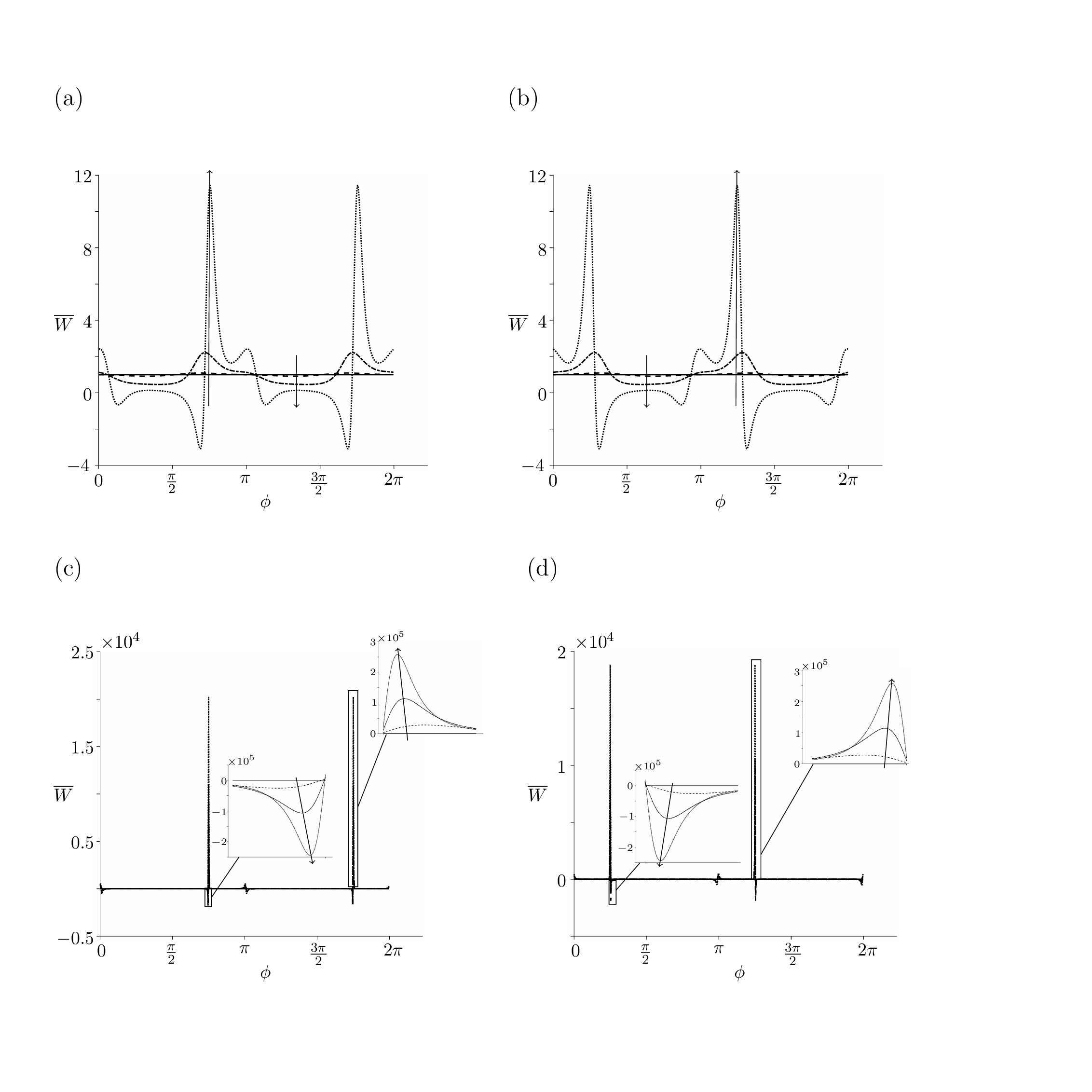}
\caption{Mean rate of working versus the initial orientation $\phi$, where $\mu_2=\mu_3=0$. (a) and (b) depict small positive and negative $\mu_1$ values; $0$ (solid line), $\pm 0.1$ (dashed line), $\pm 1$ (dash-dotted line) and $\pm 5$ (dotted line). (c) and (d) depict larger $\mu_1$ values; $0$ (solid line), $\pm 300$ (dashed line), $\pm 600$ (dash-dotted line), $\pm 900$ (dotted line), where the arrows denote increasing $\mu_1$.}
\label{fig:ActiveOnly}
\end{figure}

The active-only regime considers $\mu_2$ and $\mu_3$ zero with $\mu_1$ non-zero. The mean swimming velocity is considered in figure \ref{fig:ActiveOnlyU} and the mean rate of working in figure \ref{fig:ActiveOnly}. For $\mu_1=0$ we regain the Newtonian result and hence both mean swimming velocity and mean rate of working are independent of fibre angle. For non-zero active parameter $\mu_1$, the mean swimming velocity and mean rate of working vary considerably with fibre angle. In particular negative mean swimming velocity -- {\it i.e.}\ reversal of swimming direction -- and negative mean rate of working are observed in certain regimes for large $\mu_1$, with a sudden and dramatic switch in sign close to $\phi=3\pi/4$ (figures \ref{fig:ActiveOnlyU} (c), (d) and \ref{fig:ActiveOnly} (c), (d)). Note that this can be resolved through refinement of the plotting grid and is not a discontinuity. A change from `pusher' to `puller' type active behaviour (equivalent to a change in sign of $\mu_1$) is equivalent to a reflection in the line $\phi=\pi/2$ ($3\pi/2$).

\subsection{Regime (iii): Nearly-isotropic behaviour in leading order mean swimming velocity and rate of working}
\begin{figure}
\centering
\includegraphics[scale=0.6, trim=1.1cm 1.7cm 2.1cm 0.8cm,clip]{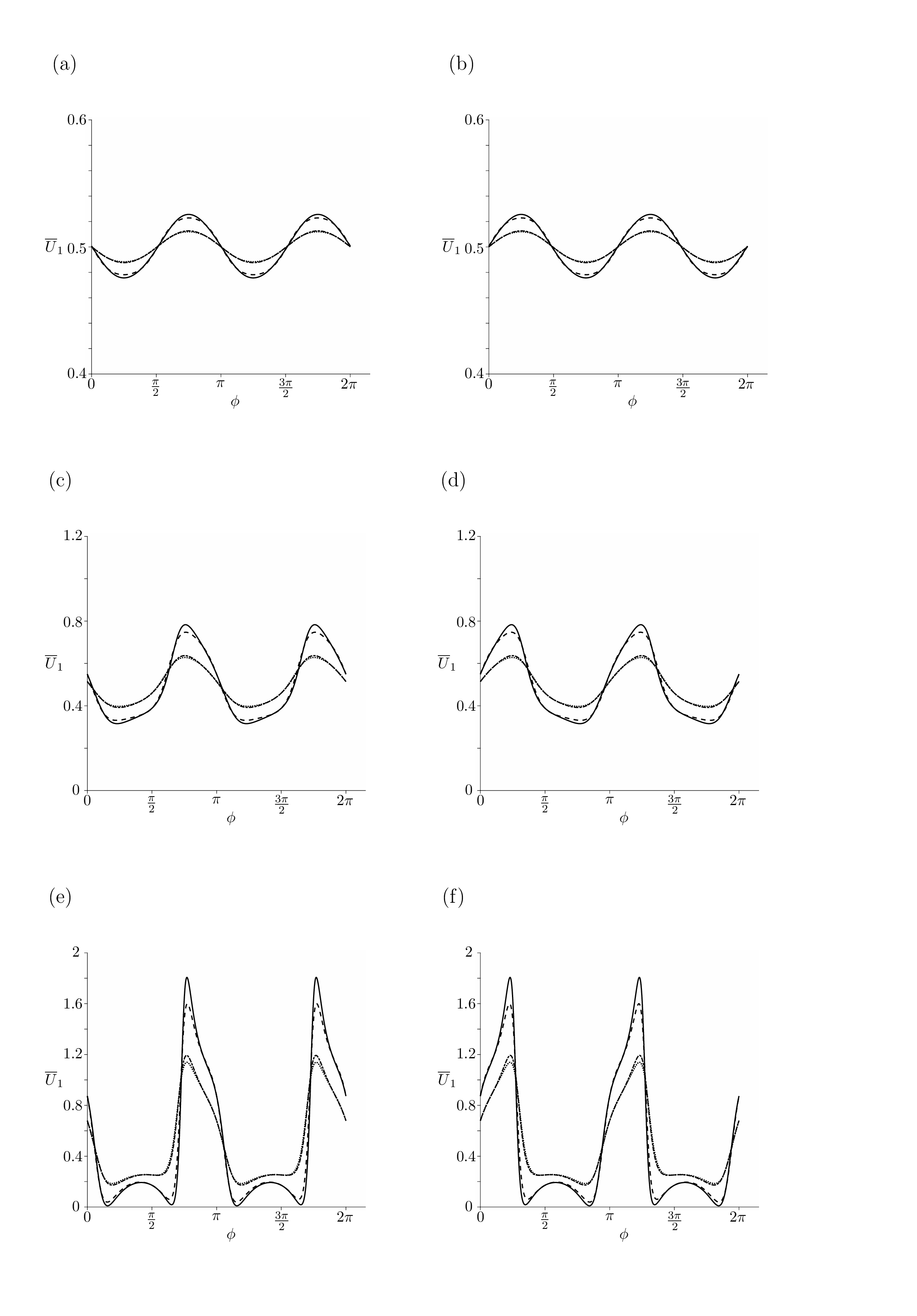}
\caption{Mean swimming velocity versus $\phi$ where parameters $\mu_1,\, \mu_2$ and $\mu_3$ take values up to $5$. (a) and (b) show $\mu_1=\pm 0.1$, (c) and (d) show $\mu_1=\pm 1$ and (e) and (f) show $\mu_1=\pm 5$. Each line depicts a different $\mu_2$ and $\mu_3$ combination; $\mu_2=0,\, \mu_3=0$ (solid line), $\mu_2=1,\, \mu_3=0$ (dashed line), $\mu_2=0,\, \mu_3=1$ (dash-dotted line) and $\mu_2=1,\, \mu_3=1$ (dotted line).}
\label{fig:NearlyIsoU}
\end{figure}

\begin{figure}
\centering
\includegraphics[scale=0.6, trim=1cm 1cm 2cm 0.3cm,clip]{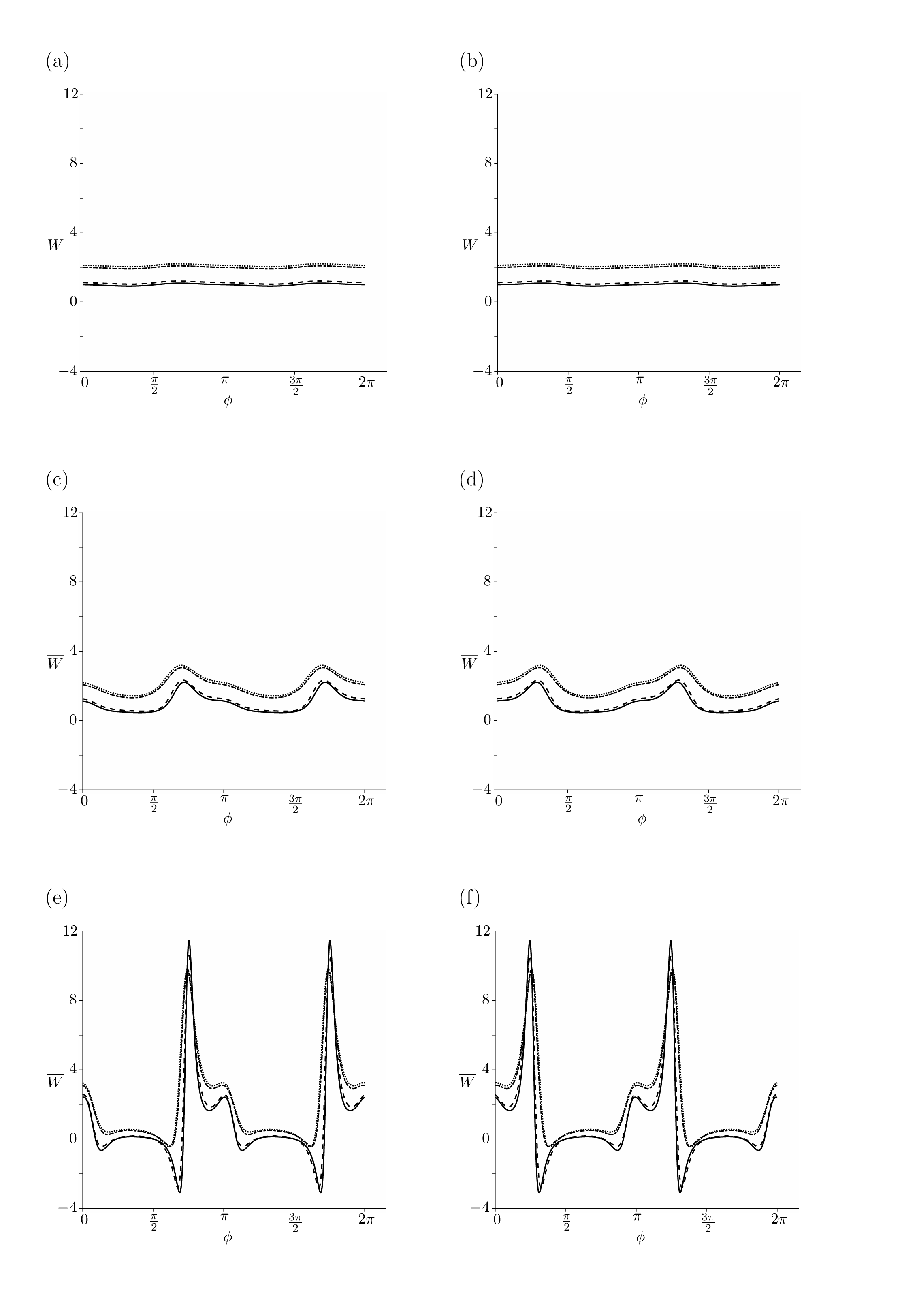}
\caption{Mean rate of working versus $\phi$ where parameters $\mu_1,\, \mu_2$ and $\mu_3$ take values up to $5$. (a) and (b) show $\mu_1=\pm 0.1$, (c) and (d) show $\mu_1=\pm 1$ and (e) and (f) show $\mu_1=\pm 5$. Each line depicts a different $\mu_2$ and $\mu_3$ combination; $\mu_2=0,\, \mu_3=0$ (solid line), $\mu_2=1,\, \mu_3=0$ (dashed line), $\mu_2=0,\, \mu_3=1$ (dash-dotted line) and $\mu_2=1,\, \mu_3=1$ (dotted line).}
\label{fig:NearlyIso}
\end{figure}

A small perturbation away from the isotropic case is considered here; $\mu_1,\, \mu_2$ and $\mu_3$ take values up to $5$. When $\mu_1$ is much smaller than one and positive (figures \ref{fig:NearlyIsoU} (a) and (b)) a small perturbation away from the Newtonian case is observed. As $\mu_1$ continues to increase, angular dependence becomes more prevalent. For the mean swimming velocity, $\mu_2$ has minimal impact, while $\mu_3$ reduces the range of values the background flow can take. For the mean rate of working (figure \ref{fig:NearlyIso}), $\mu_2$ again has little impact on the results and the effect of increasing $\mu_3$ is to increase the cost of swimming. When $\mu_1=\pm 5$ (figures \ref{fig:NearlyIso} (e) and (f)), the mean rate of working may become negative and the effect of increasing $\mu_3$ is to reduce the range of values the mean rate of working will take.

\subsection{Regime (iv): The effect of large rheological parameters on leading order mean swimming velocity and rate of working}

\begin{figure}
\centering
\includegraphics[scale=0.6, trim=1.1cm 1.7cm 2.1cm 0.8cm,clip]{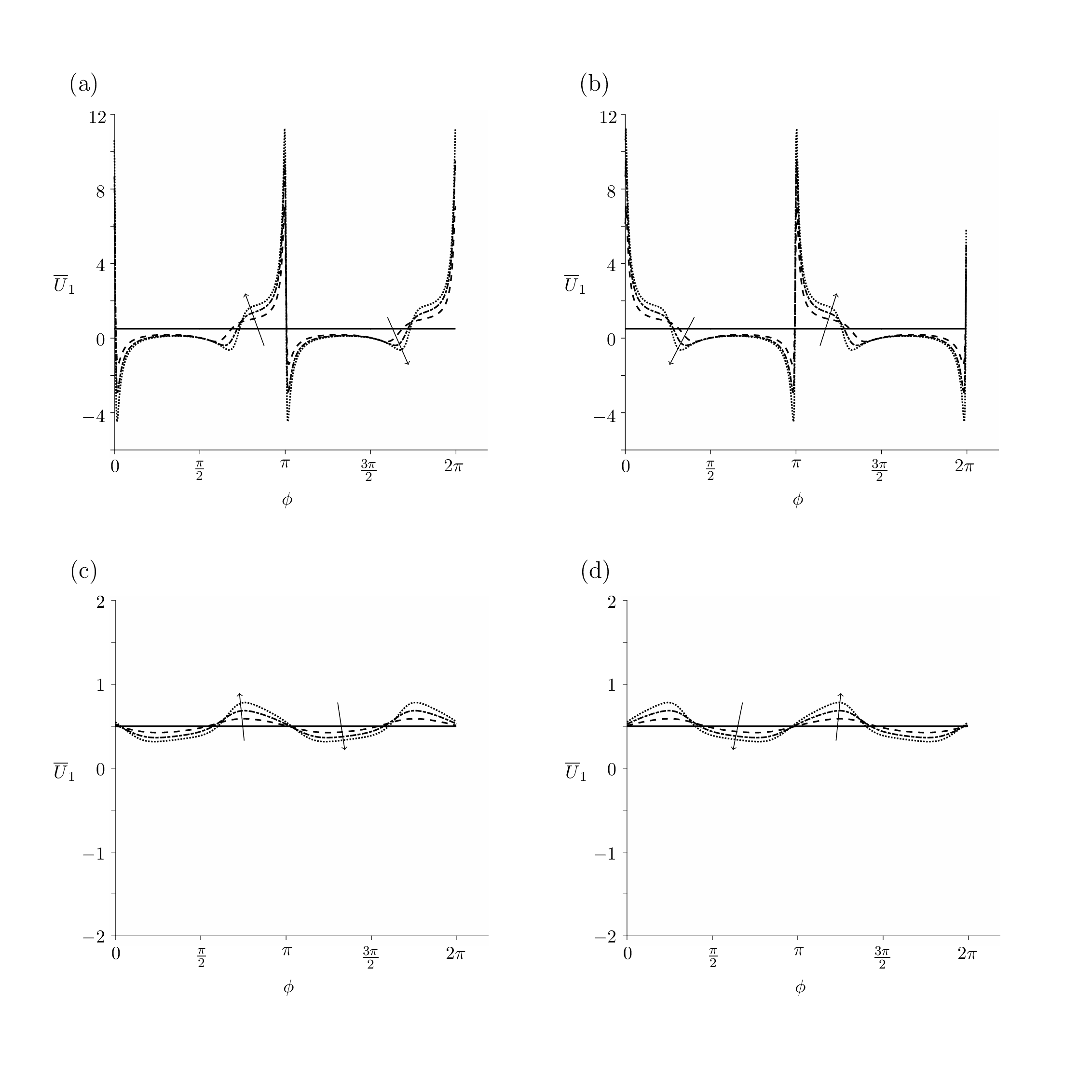}
\caption{Mean swimming velocity versus $\phi$ where at least one of $\mu_1,\, \mu_2$ and $\mu_3$ are much larger than one. (a) and (c) depict positive $\mu_1$ values and (b) and (d) depict negative $\mu_1$ values. The values $\mu_1$ takes are $0$ (solid line), $\pm 300$ (dashed line), $\pm 600$ (dash-dotted line) and $\pm 900$ (dotted line). In (a) and (b), $\mu_2=900,\, \mu_3=0$ and in (c) and (d), $\mu_2=0,\, \mu_3=900$, where the arrows denote increasing $\mu_1$.}
\label{fig:LargeParamsU}
\end{figure}

\begin{figure}
\centering
\includegraphics[scale=0.6, trim=1cm 1cm 2cm 0.3cm,clip]{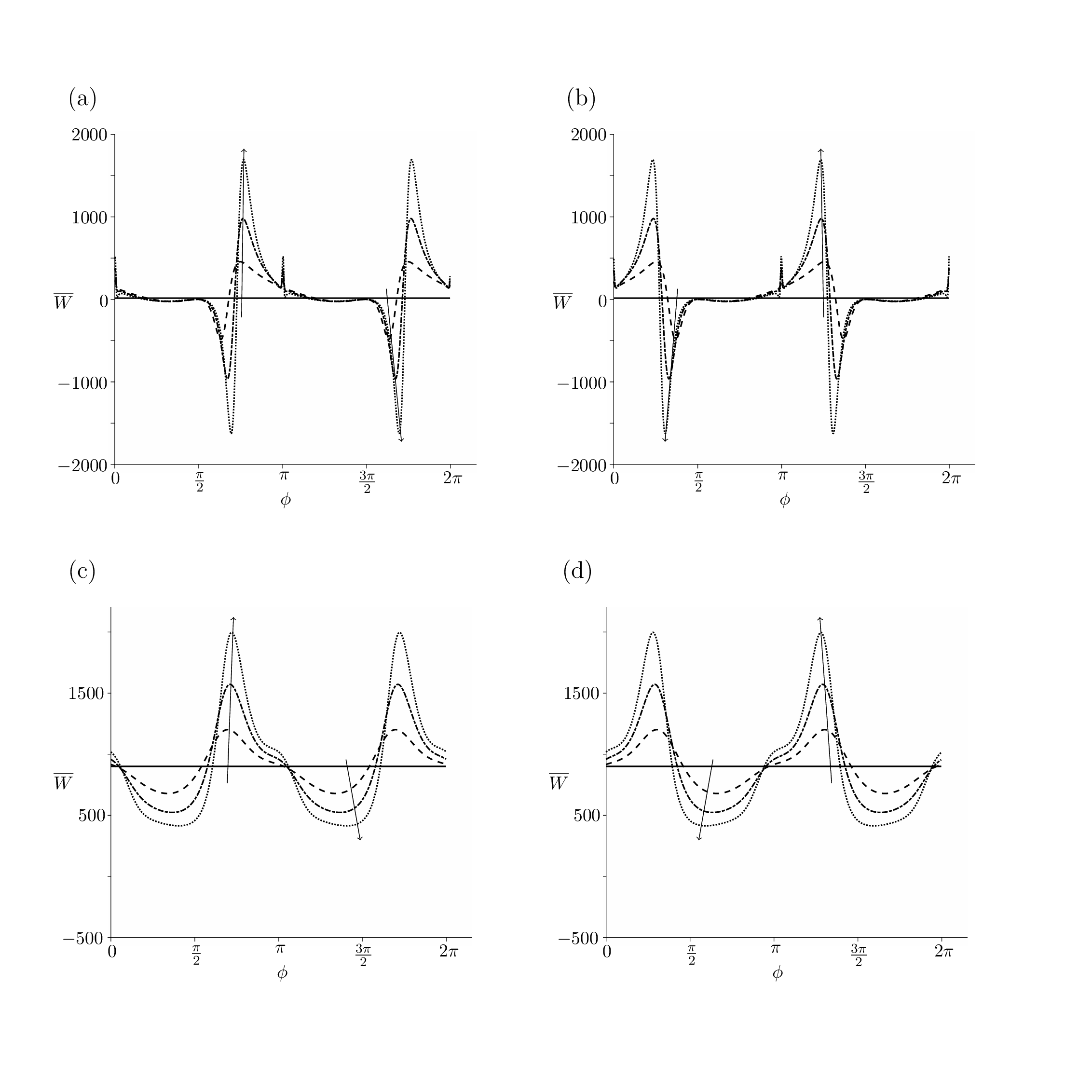}
\caption{Mean rate of working versus $\phi$ where at least one of $\mu_1,\, \mu_2$ and $\mu_3$ are much larger than one. (a) and (c) depict positive $\mu_1$ values and (b) and (d) depict negative $\mu_1$ values. The values $\mu_1$ takes are $0$ (solid line), $\pm 300$ (dashed line), $\pm 600$ (dash-dotted line) and $\pm 900$ (dotted line). In (a) and (b), $\mu_2=900,\, \mu_3=0$ and in (c) and (d), $\mu_2=0,\, \mu_3=900$, where the arrows denote increasing $\mu_1$.}
\label{fig:LargeParams}
\end{figure}

The final regime is where at least one of $\mu_1,\, \mu_2$ and $\mu_3$ are much larger than one. Figures \ref{fig:LargeParamsU} and  \ref{fig:LargeParams} depict how the mean swimming velocity and mean rate of working change with initial orientation angle, $\phi$. When either $\mu_2$ or $\mu_3$ are non-zero, the steep peaks which occurred at $\phi=3\pi/4\,(7\pi/4)$ (figure \ref{fig:ActiveOnlyU} (c)) and at $\phi=\pi/4\,(5\pi/4)$ (figure \ref{fig:ActiveOnlyU} (d)) within regime (ii) no longer appear. Further, when $\mu_2$ is non-zero and $\mu_3=0$, the mean swimming velocity becomes negative for certain initial orientation angles, {\it i.e.} the swimming direction is reversed. When $\mu_3$ becomes non-zero, the results collapse down towards the Newtonian case, altered predominantly by the active parameter, $\mu_1$. Similar results are seen for the mean rate of working, however for non-zero $\mu_3$ (figure \ref{fig:LargeParams} (c) and (d)) the reference value about which variations occur is significantly increased.

\subsection{Orientation, velocity and stream function}

\begin{figure}
\centering
\includegraphics[scale=0.63, trim=0.95cm 1.5cm 0.5cm 1cm,clip]{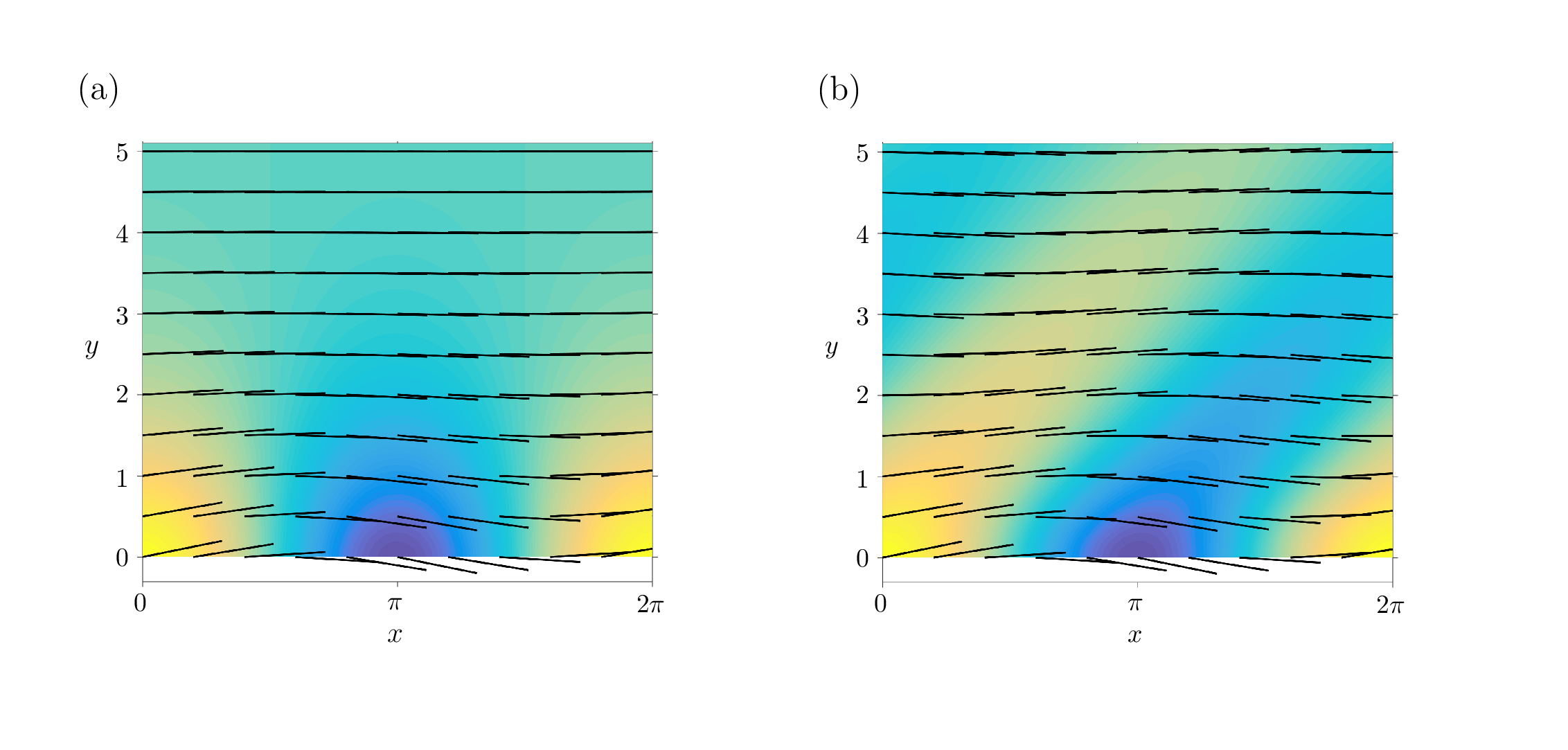}
\caption{Fibre angle, $\phi+\varepsilon\theta_0$, in passive and active regimes: (a) the passive regime ($\mu_1=0,\, \mu_2=\mu_3=5$) and (b) the active-only regime ($\mu_1=5,\, \mu_2=\mu_3=0$). In each graph $t=0$, $\varepsilon=0.2$ and the initial orientation angle is $\phi=0$. See movie 1.}
\label{fig:Angle}
\end{figure}

\begin{figure}
\centering
\includegraphics[scale=0.65, trim=1.19cm 1cm 0cm 0.6cm,clip]{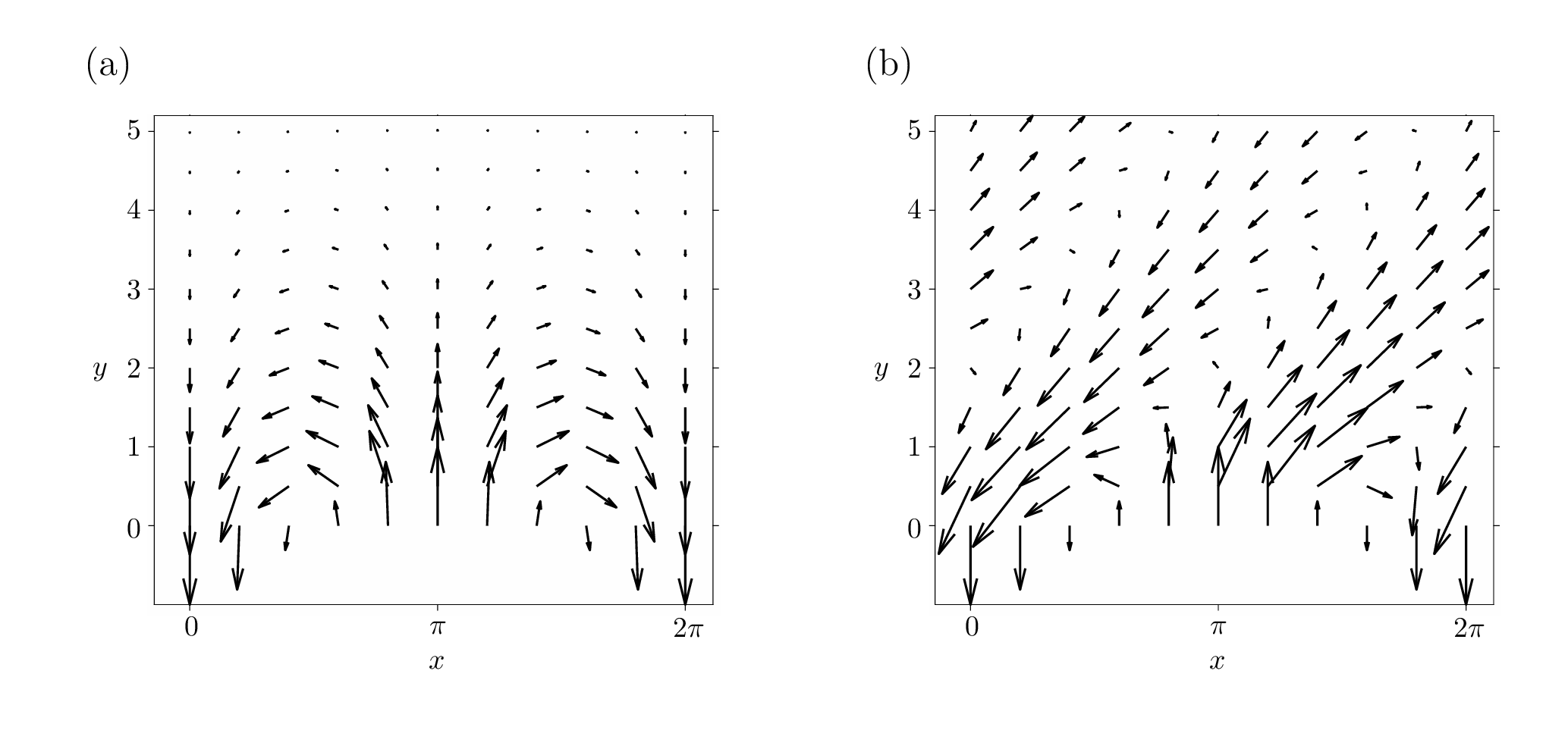}
\caption{Velocity field in passive and active regimes: (a) passive regime ($\mu_1=0,\, \mu_2=\mu_3=5$) and (b) active-only regime ($\mu_1=5,\, \mu_2=\mu_3=0$). In each graph $t=0$ and the initial orientation angle is $\phi=0$. See movie 2.}
\label{fig:Vel}
\end{figure}

\begin{figure}
\centering
\includegraphics[scale=0.63, trim=1.1cm 2.3cm 2cm 1.6cm,clip]{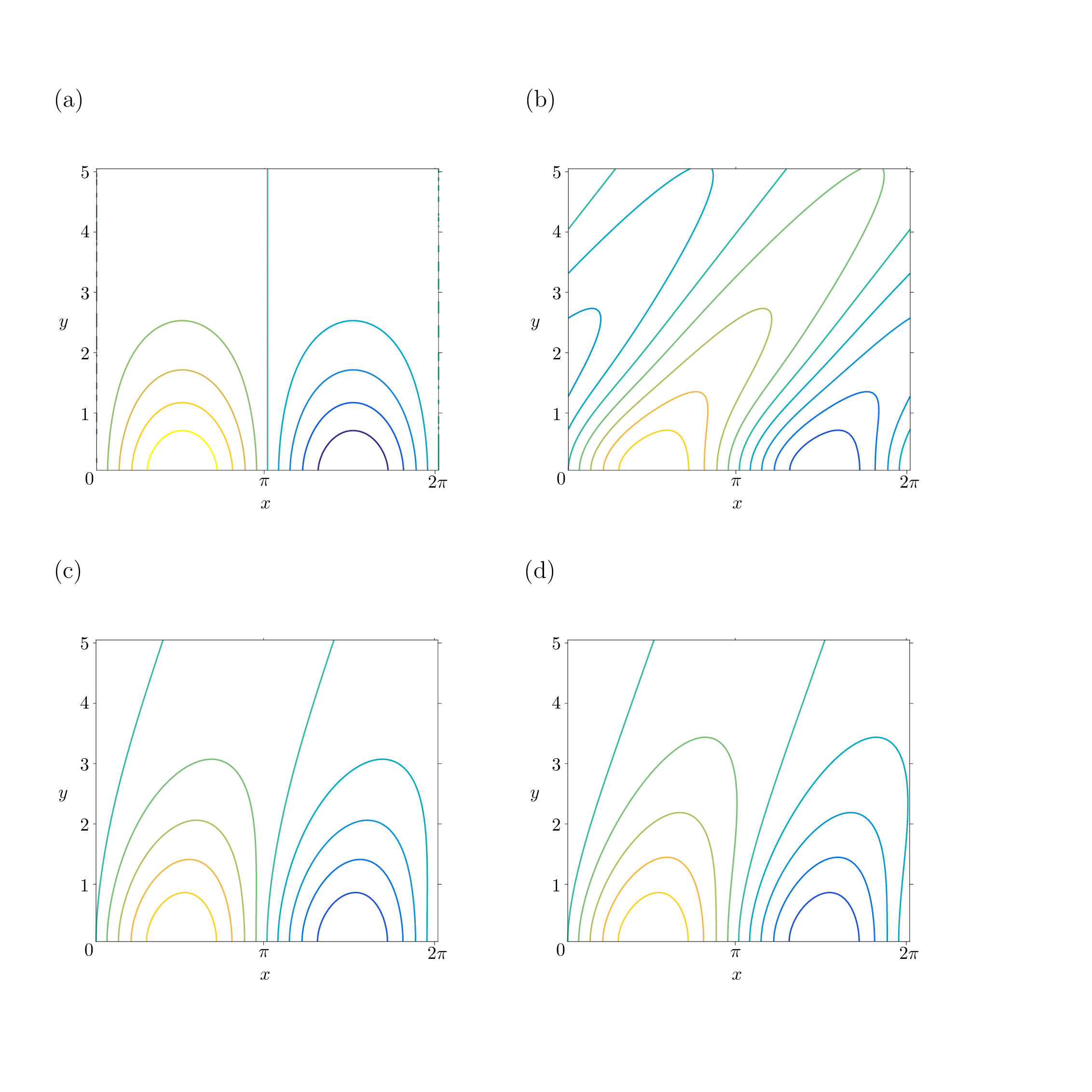}
\caption{Instantaneous streamlines in the four regimes considered: (a) passive regime ($\mu_1=0,\, \mu_2=\mu_3=5$), (b) active-only regime ($\mu_1=5,\, \mu_2=\mu_3=0$), (c) nearly-isotropic regime ($\mu_1=\mu_3=1,\, \mu_2=0$) and (d) regime where at least one of the parameters is much larger than one ($\mu_1=\mu_2=\mu_3=900$). In each graph $t=0$ and the initial orientation angle is $\phi=0$. See movie 3.}
\label{fig:Psi}
\end{figure}

Finally, to understand how fibre orientation and velocity are impacted by the anisotropic fluid properties, the orientation angle ($\phi+\theta$) and velocity are considered in an active and passive regime, and stream function are considered in all four regimes of interest (figures \ref{fig:Angle}, \ref{fig:Vel} and \ref{fig:Psi} respectively). Each variable is plotted for one wavelength of the sheet, $x=0$ to $2\pi$. We focus on the case where the fibres are aligned with the sheet, {\it i.e.} $\phi=0$ and plot results at time $t=0$ ({\it i.e.} the start of one oscillation period). See also Supplementary Movies.

Considering first the fibre orientation, in all cases perturbations to the initial orientation angle are greater in the vicinity of the sheet and are displaced with the movement of the sheet (figure \ref{fig:Angle} and movie 1).
For passive rheology, the fibre reorientation is dampened very quickly moving away from the sheet (figure \ref{fig:Angle} (a)).
Once $\mu_1$ is non-zero, fibre displacement appears further away from the sheet (figure \ref{fig:Angle} (b)) and movements propagate to the right.


Figure \ref{fig:Vel} and movie 2 show the velocity components in $x$ and $y$.
The leading-order velocity decays quickly moving away from the sheet, evident in figure \ref{fig:Vel} (a); in the active-only regime (figure \ref{fig:Vel} (b)) the flow decays more slowly.
 The velocity field shows a similar rightward propagation to fibre angle in the active fluid case (figures \ref{fig:Angle} (b) and \ref{fig:Vel} (b)).
  These results are mirrored in figure \ref{fig:Psi} and movie 3, where the streamlines of the resulting flow are displayed.
  In the passive regime (i), the streamlines are symmetric about $x=\pi$ with anticlockwise flow between $x=0$ and $\pi$ and clockwise flow for $x=\pi$ to $2\pi$. Introducing $\mu_1$ distorts the streamlines and, when $\mu_2=\mu_3=0$, the streamlines are deflected to the right (figure \ref{fig:Psi}(b)); introducing the other two parameters dampens this deflection (figures \ref{fig:Psi} (c) and (d)).

\section{Discussion} \label{sec:conc}


The classical Taylor's swimming sheet problem was modified to account for transverse isotropy, modelling swimming in fibre-reinforced fluids or active media. Quantities of interest were the steady background flow, which corresponds to the mean swimming velocity, and mean rate of working. The results presented were non-dimensional. 
The dimensional velocity scales with the wave speed, and the rate of working scales with the square of the frequency, the viscosity and the wavenumber. The ratio of the mean swimming speed to the wave speed is proportional to $1/\varepsilon^2$. When $\overline{U}_1$ takes the maximum value found here, such that $\overline{U}_1\approx 40$, this corresponds to swimming faster than the wave speed when $\varepsilon>1/\sqrt{\overline{U}_1}\approx 0.16$.
Note that swimming with a prescribed beat amplitude and frequency, regardless of the rheology of the fluid, will not in general be achievable in a real biological system.

When $\mu_1$ is zero, no dependence on initial orientation angle is observed and the mean swimming velocity takes on the Newtonian value, regardless of the size of the other parameters. For non-zero $\mu_1$ certain initial orientation angles enable less energetically costly but slower swimming, with lower mean rates of working and swimming velocities. The initial angle maximising the mean swimming velocity and the mean rate of working are not in general the same. When active fibres are parallel to the swimming direction, swimming is both faster and more energetically costly compared to active fibres perpendicular to the swimming direction. The sheet utilises the surrounding environment to boost its velocity, a result seen in \citet{leshansky2009enhanced}, for point-like obstacles, and \citet{chrispell2013actuated}, for swimming in viscoelastic fluids near walls. However, these authors also predict an increased swimming efficiency, a result not seen here.
A change from `pusher' to `puller' behaviour (equivalent to a change in sign of $\mu_1$) equates to a reflection of the initial fibre angle in the $y$-axis. The activity parameter $\mu_1$ allows the mean rate of working to take on negative values, suggesting that the tension/stresslet exerted by the fibres on the sheet may at times overcome the work the sheet does on the fluid to move. For some values of $\mu_1$ the mean swimming velocity may be negative, indicating a reversal of swimming direction; this change is dependent on the uniform orientation angle $\phi$, a result also observed for rotated viscoelastic networks \citep{wrobel2016enhanced}. The inclusion of active behaviour dramatically changes the streamlines and flow field.

For a passive transversely-isotropic fluid, {\it i.e.} $\mu_1=0$, increasing the magnitude of the viscosity-like parameters increases the work the sheet must do on the fluid to swim; the fluid becomes more difficult to move through. The mean rate of working was found to be approximately linear in the parallel viscosity $\mu_\parallel=1+(\mu_2+4\mu_3)/2$, with a small additional dependence on $\mu_3$. For an active isotropic fluid ($\mu_1$ is non-zero and $\mu_2=\mu_3=0$) we observe potentially unphysical behaviour when $\mu_1$ is increased sufficiently, with rapid large variations with respect to $\phi$ in both mean swimming velocity and rate of working. Note however that, a large value of $\mu_1$ with $\mu_2=\mu_3=0$ may not represent a physically realistic fluid. These unphysical effects are reduced by the inclusion of $\mu_3$ in particular, and to a lesser extent $\mu_2$, damping these large variations.

When the rheological parameters are all non-zero, increasing the anisotropic shear viscosity causes the mean swimming velocity to collapse down towards the Newtonian result, altered slightly by the active properties of the fluid. However the mean rate of working is increased in general. The anisotropic extensional viscosity has a similar but much smaller effect.

This study demonstrates that locomotion in active fluids is dramatically different from locomotion in passive fluids.
Our finding of zero, and indeed negative rate of working for some angular configurations and sufficiently large values of the active parameter $\mu_1$ is a consequence of the energy input to the system by active stress. This phenomenon may be related to superfluid behaviour recently observed in bacteria suspensions \citep{lopez2015turning}. Further, these results are suggestive that active stresses in the bulk may enhance the motion of individual swimmers. It has long been known that some flagellated swimmers may propel more rapidly in polymeric than Newtonian fluids \citep{schneider1974effect}. An increase in the anisotropy of the slender body drag coefficients has been proposed as one underlying mechanism \citep{berg1979movement, magariyama2002mathematical}; recently \citet{martinez2014flagellated} demonstrated that viscosity reduction associated with high speed flagellar rotation provides an alternative explanation. The present model does not support a change to mean swimming speed purely through fluid anisotropy; because we analysed only 2D propulsion with constant viscosity-like parameters we are unable to comment on the effect of shear-thinning on rotation.

The passive region of parameter space with $\mu_1=0$ represents the anisotropic characteristics of the aligned passive microstructure of cervical mucus. Key aspects which may be explored in future work include shear-dependent viscosity and dispersion of fibre alignment. The active regime $\mu_1\neq 0$ may be considered as a model of motility through an active aligned medium, which may capture some of the essential physics of sperm migration through ciliated epithelium in the female reproductive tract. Our predictions could be tested experimentally by constructing an actuated membrane and studying the dynamics of an overlying suspension of swimming bacteria or microrods.

This study has opened up a number of exciting future research directions. These include (but are not limited to) investigating the effects of viscoelasticity ({\it cf.}~\citeauthor{kruse2005generic}, \citeyear{kruse2005generic}), fibre dispersion ({\it cf.}~\citeauthor{woodhouse2012spontaneous}, \citeyear{woodhouse2012spontaneous}) and the presence of walls ({\it cf.}~\citeauthor{katz1974propulsion}, \citeyear{katz1974propulsion}). Similarly, coupling the flagellar elastic behaviour to the viscous fluid mechanics to determine the effect on the beat pattern (\citeauthor{riley2014}, \citeyear{riley2014}), and a full 3D computational treatment of the problem would be of interest. The model may also be developed to apply to peristaltic pumping by taking into account a cylindrical tube geometry. Taylor's swimming sheet has inspired decades of research into biological propulsion; the study presented here shows that Taylor's model continues to enable insight into novel areas of active fluid mechanics.


\section*{Acknowledgements}

GC is supported by a Biotechnology and Biological Sciences Research Council (BBSRC) Industrial CASE Studentship (BB/L015587/1). The authors acknowledge Chloe Spalding and Alex Tisbury for their contributions to an initial student project leading to this problem, and Craig Holloway, Dr Meurig Gallagher and Dr Matt Hicks for valuable discussions.

\begin{appendices}
\section{Components of the stress tensor}\label{appA}
Assuming that the velocity takes the form $\boldsymbol{u}=(u(x,y,t),v(x,y,t))$, the components of the stress tensor are calculated as,
\begingroup
\addtolength{\jot}{0.7em}
\begin{eqnarray}
&\sigma_{11}&=-p+\mu_1(\cos\phi-\theta\sin\phi)^2+\left(2+\mu_2(\cos\phi-\theta\sin\phi)^4 \right. \nonumber \\
& & \left. \quad +4\mu_3(\cos\phi-\theta\sin\phi)^2\right)\dpd{u}{x}+\biggl( \mu_2(\cos\phi-\theta\sin\phi)^3(\sin\phi+\theta\cos\phi) \nonumber \\
& & \quad +2\mu_3(\cos\phi-\theta\sin\phi)(\sin\phi+\theta\cos\phi) \biggr) \left(\dpd{u}{y}+\dpd{v}{y}\right) \nonumber \\
&\quad &+\mu_2(\cos\phi-\theta\sin\phi)^2(\sin\phi+\theta\cos\phi)^2\dpd{v}{y},
\end{eqnarray}
\begin{eqnarray}
&\sigma_{12}&=\sigma_{21}\nonumber \\
&\quad & =\mu_1(\cos\phi-\theta\sin\phi)(\sin\phi+\theta\cos\phi) \nonumber \\
 &\quad & +\left(1+\mu_2(\cos\phi-\theta\sin\phi)^2(\sin\phi+\theta\cos\phi)^2+\mu_3(1-\theta^2)\right)\left(\dpd{u}{y}+\dpd{v}{x}\right)\nonumber \\
&\quad &+\mu_2(\cos\phi-\theta\sin\phi)^3(\sin\phi+\theta\cos\phi)\dpd{u}{x}\nonumber \\
&\quad &+\mu_2(\cos\phi-\theta\sin\phi)(\sin\phi+\theta\cos\phi)^3\dpd{v}{y},
\end{eqnarray}
\begin{eqnarray}
&\sigma_{22}&=-p+\mu_1(\sin\phi+\theta\cos\phi)^2+\left(2+\mu_2(\sin\phi+\theta\cos\phi)^4\right.\nonumber \\
&\quad &\left. +4\mu_3(\sin\phi+\theta\cos\phi)^2\right)\dpd{v}{y}+\biggl(\mu_2(\cos\phi-\theta\sin\phi)(\sin\phi+\theta\cos\phi)^3\nonumber \\
&\quad & 2\mu_3(\cos\phi-\theta\sin\phi)(\sin\phi+\theta\cos\phi)\biggr) \left(\dpd{u}{y}+\dpd{v}{x}\right)\nonumber \\
&\quad &+\mu_2(\cos\phi-\theta\sin\phi)^2(\sin\phi+\theta\cos\phi)^2\dpd{u}{x}.
\end{eqnarray}
\endgroup

\section{}\label{appB}
The components of the matrix $\mbi{L}$ are calculated as
\begin{eqnarray}
{L}_{11}&=&(1+\frac{\mu_2}{4}\sin^22\phi+\mu_3)(\lambda^4-2\lambda^2+1)-\mu_2\lambda^2\cos4\phi\nonumber \\
& &\qquad +\mu_1\left[2\sin2\phi(\lambda\cos^2\phi-\lambda^3\sin^2\phi)
 +(\lambda+\lambda^3)\cos2\phi\sin2\phi\right], \\
{L}_{12}&=&-\mu_1\left[2\lambda^2\sin^22\phi+\cos2\phi((\lambda^2+\lambda^4)\sin^2\phi-(1+\lambda^2)\cos^2\phi)\right] \nonumber \\
& &\qquad +\frac{\mu_2}{2}(\lambda^3+\lambda)\sin4\phi,
\end{eqnarray}
where ${L}_{22}={L}_{11}$ and ${L}_{21}=-{L}_{12}$.

\section{}\label{appC}
The balance of \eqref{FullGov} at order $\varepsilon^2$ is given by
\begin{eqnarray}
(1&+&\frac{\mu_2}{4}\sin^22\phi+\mu_3)\nabla^4\psi_1-\mu_1\left(2\sin2\phi\dmd{\theta_1}{2}{x}{}{y}{}+\cos2\phi\left(\dpd[2]{\theta_1}{x}-\dpd[2]{\theta_1}{y}\right)\right) \nonumber \\
&+& \mu_2\left(\cos4\phi\dmd{\psi_1}{4}{x}{2}{y}{2}+\frac{\sin4\phi}{2}\left(\dmd{\psi_1}{4}{x}{}{y}{3}-\dmd{\psi_1}{4}{x}{3}{y}{}\right)\right)\nonumber \\
&-&\mu_1\left[2\sin2\phi\left(\theta_0\left(\dpd[2]{\theta_0}{y}-\dpd[2]{\theta_0}{x}\right)+\left(\dpd{\theta_0}{y}\right)^2-\left(\dpd{\theta_0}{x}\right)^2\right)\right. \nonumber \\
& & \hspace{1.3cm}\left.+\cos2\phi\left(4\left(\dpd{\theta_0}{x}\dpd{\theta_0}{y}+\theta_0\dmd{\theta_0}{2}{x}{}{y}{}\right)\right)\right] -\mu_2\left[\sin4\phi\left(2\dmd{\theta_0}{2}{x}{}{y}{}\dmd{\psi_0}{2}{x}{}{y}{} \right.\right. \nonumber \\
& &\left.\left. -\frac{\theta_0}{2}\left(\dpd[4]{\psi_0}{x}-3\dmd{\psi_0}{4}{x}{2}{y}{2}+\dpd[4]{\psi_0}{y}\right) +\frac{1}{2}\left(\dpd[2]{\theta_0}{x}-\dpd[2]{\theta_0}{y}\right)\left(\dpd[2]{\psi_0}{y}-\dpd[2]{\psi_0}{x}\right)\right.\right. \nonumber \\
& & \qquad \left.\left.-\dpd{\theta_0}{y}\left(\dpd[3]{\psi_0}{y}-3\dmd{\psi_0}{3}{x}{2}{y}{}\right)+\dpd{\theta_0}{x}\left(3\dmd{\psi_0}{3}{x}{}{y}{2}-\dpd[3]{\psi_0}{x}\right)+\theta_0\dmd{\psi_0}{4}{x}{2}{y}{2}\right)\right. \nonumber \\
& &\qquad \left.+\cos4\phi\left(2\theta_0\left(\dmd{\psi_0}{4}{x}{3}{y}{}-\dmd{\psi_0}{4}{x}{}{y}{3}\right)+\left(\dpd[2]{\theta_0}{x}-\dpd[2]{\theta_0}{y}\right)\dmd{\psi_0}{2}{x}{}{y}{} \right.\right. \nonumber \\
& & \qquad \left.\left. -\dpd{\theta_0}{x}\left(\dpd[3]{\psi_0}{y}-3\dmd{\psi_0}{3}{x}{2}{y}{}\right)-\dpd{\theta_0}{y}\left(3\dmd{\psi_0}{3}{x}{}{y}{2}-\dpd[3]{\psi_0}{x}\right)\right.\right. \nonumber \\
& & \hspace{5cm} \left.\left. -\dmd{\theta_0}{2}{x}{}{y}{}\left(\dpd[2]{\psi_0}{y}-\dpd[2]{\psi_0}{x}\right)\right)\right]=0.
\end{eqnarray}

\end{appendices}

\end{document}